\pdfoutput=1

\documentclass[12pt,a4paper]{article}

\usepackage{ifthen} 
\newboolean{pdflatex}
\setboolean{pdflatex}{true} 

\newboolean{articletitles}
\setboolean{articletitles}{true} 

\newboolean{uprightparticles}
\setboolean{uprightparticles}{false} 

\newboolean{inbibliography}
\setboolean{inbibliography}{false} 

\usepackage{microtype}
\usepackage{lineno}  
\usepackage{xspace} 
\usepackage{caption} 

\usepackage{graphicx}  
\usepackage{color}
\usepackage{colortbl}
\graphicspath{{./figs/}} 

\usepackage{amsmath} 
\usepackage{amssymb}
\usepackage{amsfonts}
\usepackage{upgreek} 

\newcommand*\patchAmsMathEnvironmentForLineno[1]{%
\expandafter\let\csname old#1\expandafter\endcsname\csname #1\endcsname
\expandafter\let\csname oldend#1\expandafter\endcsname\csname
end#1\endcsname
 \renewenvironment{#1}%
   {\linenomath\csname old#1\endcsname}%
   {\csname oldend#1\endcsname\endlinenomath}%
}
\newcommand*\patchBothAmsMathEnvironmentsForLineno[1]{%
  \patchAmsMathEnvironmentForLineno{#1}%
  \patchAmsMathEnvironmentForLineno{#1*}%
}
\AtBeginDocument{%
\patchBothAmsMathEnvironmentsForLineno{equation}%
\patchBothAmsMathEnvironmentsForLineno{align}%
\patchBothAmsMathEnvironmentsForLineno{flalign}%
\patchBothAmsMathEnvironmentsForLineno{alignat}%
\patchBothAmsMathEnvironmentsForLineno{gather}%
\patchBothAmsMathEnvironmentsForLineno{multline}%
}

\usepackage{hyperref}    
\usepackage[all]{hypcap} 

\def\lhcb {\mbox{LHCb}\xspace}

\ifthenelse{\boolean{uprightparticles}}%
{

 \def\Pmu         {\ensuremath{\upmu}\xspace}                 
 \def\Pnu         {\ensuremath{\upnu}\xspace}                 
                  
 \def\Ppi         {\ensuremath{\uppi}\xspace}

 \def\PDelta      {\ensuremath{\Delta}\xspace}                 
 \def\PXi      {\ensuremath{\Xi}\xspace}                 
 \def\PLambda      {\ensuremath{\Lambda}\xspace}                 
 \def\PSigma      {\ensuremath{\Sigma}\xspace}                 
 \def\POmega      {\ensuremath{\Omega}\xspace}                 
 \def\PUpsilon      {\ensuremath{\Upsilon}\xspace}                 
 
 \def\PB      {\ensuremath{\mathrm{B}}\xspace}                 
                  
 \def\PD      {\ensuremath{\mathrm{D}}\xspace}

 \def\PK      {\ensuremath{\mathrm{K}}\xspace}

 \def\PX      {\ensuremath{\mathrm{X}}\xspace}

 \def\Pb      {\ensuremath{\mathrm{b}}\xspace}                 
 \def\Pc      {\ensuremath{\mathrm{c}}\xspace}

 \def\Pi      {\ensuremath{\mathrm{i}}\xspace}

 \def\Pp      {\ensuremath{\mathrm{p}}\xspace}

}
{

 \def\Pmu         {\ensuremath{\mu}\xspace}                 
 \def\Pnu         {\ensuremath{\nu}\xspace}                 
                  
 \def\Ppi         {\ensuremath{\pi}\xspace}

 \mathchardef\PDelta="7101
 \mathchardef\PXi="7104
 \mathchardef\PLambda="7103
 \mathchardef\PSigma="7106
 \mathchardef\POmega="710A
 \mathchardef\PUpsilon="7107
                  
 \def\PB      {\ensuremath{B}\xspace}                 
                  
 \def\PD      {\ensuremath{D}\xspace}

 \def\PK      {\ensuremath{K}\xspace}

 \def\PX      {\ensuremath{X}\xspace}

 \def\Pb      {\ensuremath{b}\xspace}                 
 \def\Pc      {\ensuremath{c}\xspace}

 \def\Pi      {\ensuremath{i}\xspace}

 \def\Pp      {\ensuremath{p}\xspace}

}

\makeatletter
\ifcase \@ptsize \relax
  \newcommand{\miniscule}{\@setfontsize\miniscule{4}{5}}
\or
  \newcommand{\miniscule}{\@setfontsize\miniscule{5}{6}}
\or
  \newcommand{\miniscule}{\@setfontsize\miniscule{5}{6}}
\fi
\makeatother

\DeclareRobustCommand{\optbar}[1]{\shortstack{{\miniscule (\rule[.5ex]{1.25em}{.18mm})}
  \\ [-.7ex] $#1$}}

\def\mun        {{\ensuremath{\Pmu^-}}\xspace} 

\def\neub       {{\ensuremath{\overline{\Pnu}}}\xspace}

\def\neumb      {{\ensuremath{\neub_\mu}}\xspace}

\def\cquark    {{\ensuremath{\Pc}}\xspace}

\def\bquark    {{\ensuremath{\Pb}}\xspace}
\def\bquarkbar {{\ensuremath{\overline \bquark}}\xspace}

\def\pion   {{\ensuremath{\Ppi}}\xspace}

\def\pip    {{\ensuremath{\pion^+}}\xspace}
\def\pim    {{\ensuremath{\pion^-}}\xspace}
\def\pipm   {{\ensuremath{\pion^\pm}}\xspace}

\def\kaon    {{\ensuremath{\PK}}\xspace}
  \def\Kbar    {{\kern 0.2em\overline{\kern -0.2em \PK}{}}\xspace}

\def\KorKbar    {\kern 0.18em\optbar{\kern -0.18em K}{}\xspace}
\def\Kz      {{\ensuremath{\kaon^0}}\xspace}
\def\Kzb     {{\ensuremath{\Kbar{}^0}}\xspace}
\def\Kp      {{\ensuremath{\kaon^+}}\xspace}
\def\Km      {{\ensuremath{\kaon^-}}\xspace}
\def\Kpm     {{\ensuremath{\kaon^\pm}}\xspace}

\def\KS      {{\ensuremath{\kaon^0_{\rm\scriptscriptstyle S}}}\xspace}
\def\KL      {{\ensuremath{\kaon^0_{\rm\scriptscriptstyle L}}}\xspace}

  \def\Dbar    {{\kern 0.2em\overline{\kern -0.2em \PD}{}}\xspace}
\def\D       {{\ensuremath{\PD}}\xspace}
\def\Db      {{\ensuremath{\Dbar}}\xspace}
\def\DorDbar    {\kern 0.18em\optbar{\kern -0.18em D}{}\xspace}
\def\Dz      {{\ensuremath{\D^0}}\xspace}
\def\Dzb     {{\ensuremath{\Dbar{}^0}}\xspace}
\def\Dp      {{\ensuremath{\D^+}}\xspace}
\def\Dm      {{\ensuremath{\D^-}}\xspace}

\def\Dstarp  {{\ensuremath{\D^{*+}}}\xspace}

\def\B       {{\ensuremath{\PB}}\xspace}
\def\Bbar    {{\ensuremath{\kern 0.18em\overline{\kern -0.18em \PB}{}}}\xspace}

\def\BorBbar    {\kern 0.18em\optbar{\kern -0.18em B}{}\xspace}
\def\Bz      {{\ensuremath{\B^0}}\xspace}

\def\Bu      {{\ensuremath{\B^+}}\xspace}

\def\Bp      {{\ensuremath{\Bu}}\xspace}

\def\Bd      {{\ensuremath{\B^0}}\xspace}

  \def\Y#1S{\ensuremath{\PUpsilon{(#1S)}}\xspace}

\def\Lz          {{\ensuremath{\PLambda}}\xspace}
\def\Lbar        {{\ensuremath{\kern 0.1em\overline{\kern -0.1em\PLambda}}}\xspace}
\def\LorLbar    {\kern 0.18em\optbar{\kern -0.18em \PLambda}{}\xspace}

\def\Lc      {{\ensuremath{\Lz^+_\cquark}}\xspace}

\newcommand{\decay}[2]{\ensuremath{#1\!\to #2}\xspace}         

\def\to                 {\ensuremath{\rightarrow}\xspace}

\newcommand{\tauL}{{\ensuremath{\tau_{\rm L}}}\xspace}

\def\CP                {{\ensuremath{C\!P}}\xspace}

\newcommand{\mistag}{\ensuremath{\omega}\xspace}

\def\AT#1     {\ensuremath{A_{\mathrm{T}}^{#1}}\xspace}           

\def\C#1      {\ensuremath{\mathcal{C}_{#1}}\xspace}                       
\def\Cp#1     {\ensuremath{\mathcal{C}_{#1}^{'}}\xspace}                    
\def\Ceff#1   {\ensuremath{\mathcal{C}_{#1}^{\mathrm{(eff)}}}\xspace}        
\def\Cpeff#1  {\ensuremath{\mathcal{C}_{#1}^{'\mathrm{(eff)}}}\xspace}       
\def\Ope#1    {\ensuremath{\mathcal{O}_{#1}}\xspace}                       
\def\Opep#1   {\ensuremath{\mathcal{O}_{#1}^{'}}\xspace}                    

\newcommand{\ket}[1]{\ensuremath{|#1\rangle}}              

\newcommand{\tev}{\ifthenelse{\boolean{inbibliography}}{\ensuremath{~T\kern -0.05em eV}\xspace}{\ensuremath{\mathrm{\,Te\kern -0.1em V}}}\xspace}
\newcommand{\gev}{\ensuremath{\mathrm{\,Ge\kern -0.1em V}}\xspace}
\newcommand{\mev}{\ensuremath{\mathrm{\,Me\kern -0.1em V}}\xspace}
\newcommand{\kev}{\ensuremath{\mathrm{\,ke\kern -0.1em V}}\xspace}
\newcommand{\ev}{\ensuremath{\mathrm{\,e\kern -0.1em V}}\xspace}
\newcommand{\gevc}{\ensuremath{{\mathrm{\,Ge\kern -0.1em V\!/}c}}\xspace}
\newcommand{\mevc}{\ensuremath{{\mathrm{\,Me\kern -0.1em V\!/}c}}\xspace}
\newcommand{\gevcc}{\ensuremath{{\mathrm{\,Ge\kern -0.1em V\!/}c^2}}\xspace}
\newcommand{\gevgevcccc}{\ensuremath{{\mathrm{\,Ge\kern -0.1em V^2\!/}c^4}}\xspace}
\newcommand{\mevcc}{\ensuremath{{\mathrm{\,Me\kern -0.1em V\!/}c^2}}\xspace}

\def\mum  {\ensuremath{{\,\upmu\rm m}}\xspace}

\def\mbarn{\ensuremath{\rm \,mb}\xspace}

\def\invfb   {\ensuremath{\mbox{\,fb}^{-1}}\xspace}

\def\sec  {\ensuremath{\rm {\,s}}\xspace}

\newcommand{\stat}{\ensuremath{\mathrm{\,(stat)}}\xspace}
\newcommand{\syst}{\ensuremath{\mathrm{\,(syst)}}\xspace}

\def\gsim{{~\raise.15em\hbox{$>$}\kern-.85em
          \lower.35em\hbox{$\sim$}~}\xspace}
\def\lsim{{~\raise.15em\hbox{$<$}\kern-.85em
          \lower.35em\hbox{$\sim$}~}\xspace}

\newcommand{\mean}[1]{\ensuremath{\left\langle #1 \right\rangle}} 

\def\sPlot{\mbox{\em sPlot}}

\def\pt         {\mbox{$p_{\rm T}$}\xspace}

\def\degrees{\ensuremath{^{\circ}}\xspace}

\def\tell1  {TELL1\xspace}
\def\ukl1   {UKL1\xspace}

\newcommand{\DACP}{\ensuremath{\Delta A_{\CP}}\xspace}
\newcommand{\AcpKK}{\ensuremath{A_{\CP}(\Km\Kp)}\xspace}
\newcommand{\Acppipi}{\ensuremath{A_{\CP}(\pim\pip)}\xspace}
\newcommand{\ACP}{\ensuremath{A_{\CP}}\xspace}
\newcommand{\AKpi}{\ensuremath{A_{\rm raw}(\Km\pip)}\xspace}
\newcommand{\AKK}{\ensuremath{A_{\rm raw}(\Km\Kp)}\xspace}
\newcommand{\Apipi}{\ensuremath{A_{\rm raw}(\pim\pip)}\xspace}
\newcommand{\Araw}{\ensuremath{A_{\rm raw}}\xspace}
\newcommand{\ArawKK}{\ensuremath{A_{\rm raw}(\Km\Kp)}\xspace}

\newcommand{\ArawKpi}{\ensuremath{A_{\rm raw}(\Km\pip)}\xspace}
\newcommand{\ArawKpipi}{\ensuremath{A_{\rm raw}(\Km\pip\pip)}\xspace}
\newcommand{\ArawKzpi}{\ensuremath{A_{\rm raw}(\Kzb\pip)}\xspace}
\newcommand{\AP}{\ensuremath{A_P(\Bbar)}\xspace}
\newcommand{\APD}{\ensuremath{A_P(\Dp)}\xspace}
\newcommand{\AD}{\ensuremath{A_D(\mun)}\xspace}
\newcommand{\ADKpi}{\ensuremath{A_D(\Km\pip)}\xspace}
\newcommand{\ADpi}{\ensuremath{A_D(\pip)}\xspace}
\newcommand{\ADkz}{\ensuremath{A_D(\Kz)}\xspace}
\newcommand{\gevcNS}{\ensuremath{{\mathrm{Ge\kern -0.1em V\!/}c}}\xspace}

\def\fbar       {\overline f}
\def\dkpipi     {\decay{\Dp}{\Km\pip\pip}}
\def\dkk        {\decay{\Dz}{\Km\Kp}}
\def\dpipi      {\decay{\Dz}{\pim\pip}}
\def\dkpi       {\decay{\Dz}{\Km\pip}}
\def\dkzpi      {\decay{\Dp}{\Kzb\pip}}
\def\kzpipi     {\decay{\Kz}{\pip\pim}}

\def\alphaS      {{\ensuremath{\alpha_{\rm\scriptscriptstyle S}}}\xspace}
\def\alphaL      {{\ensuremath{\alpha_{\rm\scriptscriptstyle L}}}\xspace}

\def\alphaLS     {{\ensuremath{\alpha_{\rm\scriptscriptstyle L,S}}}\xspace}

\def\lambdaS     {{\ensuremath{\lambda_{\rm\scriptscriptstyle S}}}\xspace}
\def\lambdaL     {{\ensuremath{\lambda_{\rm\scriptscriptstyle L}}}\xspace}

\def\GammaS      {{\ensuremath{\Gamma_{\rm\scriptscriptstyle S}}}\xspace}
\def\GammaL      {{\ensuremath{\Gamma_{\rm\scriptscriptstyle L}}}\xspace}
\def\GammaLS     {{\ensuremath{\Gamma_{\rm\scriptscriptstyle L,S}}}\xspace}

\def\tauS        {{\ensuremath{\tau_{\rm\scriptscriptstyle S}}}\xspace}
\def\tauL        {{\ensuremath{\tau_{\rm\scriptscriptstyle L}}}\xspace}

\def\mS          {{\ensuremath{m_{\rm\scriptscriptstyle S}}}\xspace}
\def\mL          {{\ensuremath{m_{\rm\scriptscriptstyle L}}}\xspace}
\def\mLS         {{\ensuremath{m_{\rm\scriptscriptstyle L,S}}}\xspace}

\usepackage{cite} 
\usepackage{mciteplus}

\def\DACPval {+0.14}
\def\DACPstat {0.16}
\def\DACPsyst {0.08}

\def\AcpKKval {-0.06}
\def\AcpKKstat {0.15}
\def\AcpKKsyst {0.10}

\def\corrstat {0.23}
\def\corrtot  {0.28}

\def\Acppipival {-0.20}
\def\Acppipistat {0.19}
\def\Acppipisyst {0.10}

\begin{document}

\renewcommand{\thefootnote}{\fnsymbol{footnote}}
\setcounter{footnote}{1}


\begin{titlepage}
\pagenumbering{roman}

\vspace*{-1.5cm}
\centerline{\large EUROPEAN ORGANIZATION FOR NUCLEAR RESEARCH (CERN)}
\vspace*{1.5cm}
\hspace*{-0.5cm}
\begin{tabular*}{\linewidth}{lc@{\extracolsep{\fill}}r}
\ifthenelse{\boolean{pdflatex}}
{\vspace*{-2.7cm}\mbox{\!\!\!\includegraphics[width=.14\textwidth]{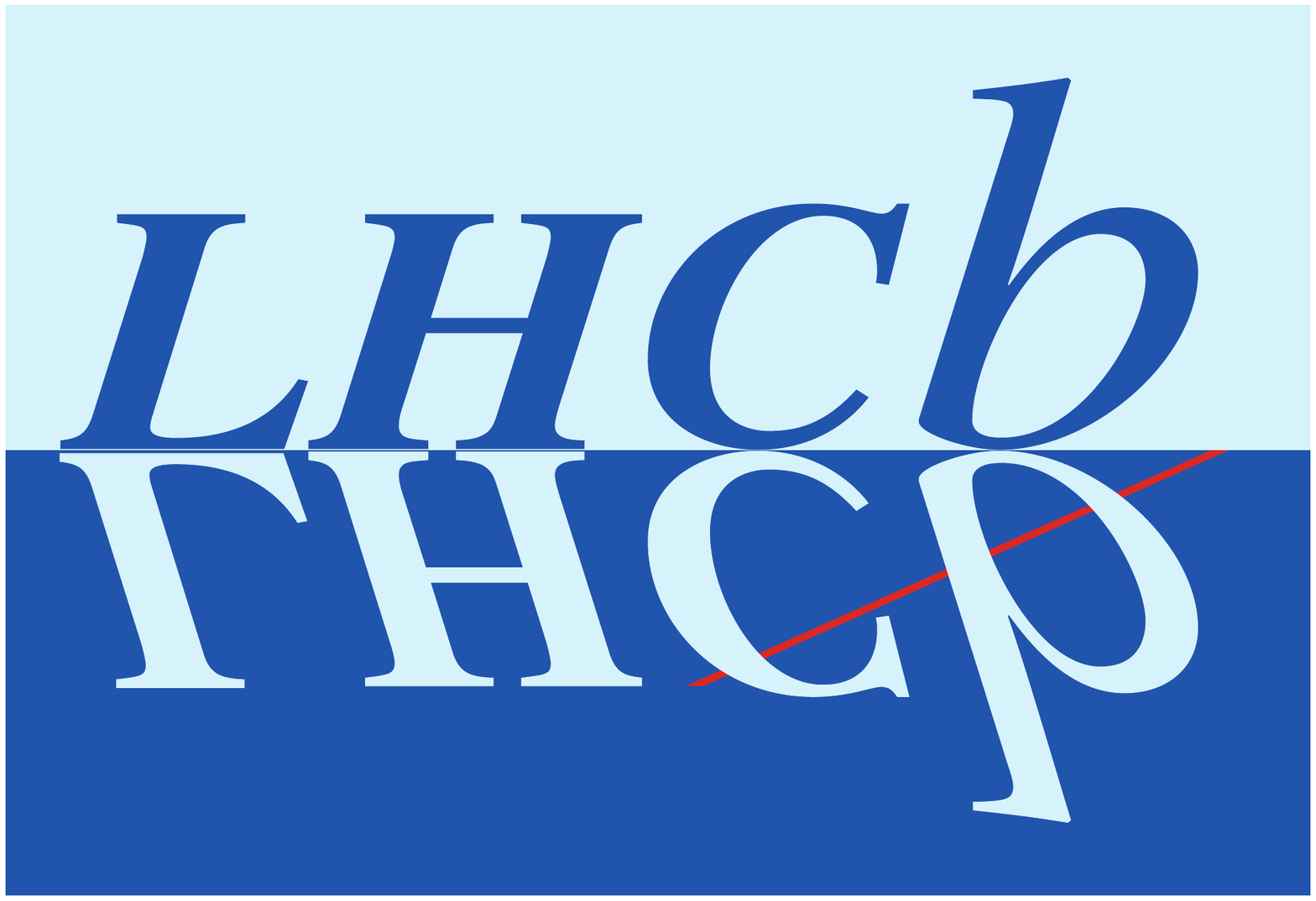}} & &}%
{\vspace*{-1.2cm}\mbox{\!\!\!\includegraphics[width=.12\textwidth]{lhcb-logo.eps}} & &}%
\\
 & & CERN-PH-EP-2014-082 \\  
 & & LHCb-PAPER-2014-013 \\  
 & & June 11, 2014 \\ 
 & & \\
\end{tabular*}

\vspace*{0.4cm}

{\bf\boldmath\huge
\begin{center}
   Measurement of $C\!P$ asymmetry in $D^0 \rightarrow K^- K^+$ and $D^0 \rightarrow \pi^- \pi^+$ decays
\end{center}
}

\vspace*{0.5cm}

\begin{center}
The LHCb collaboration\footnote{Authors are listed on the following pages.}
\end{center}

\vspace{\fill}

\begin{abstract}
  \noindent
Time-integrated $C\!P$ asymmetries in $D^0$ decays to the final states $K^- K^+$
and $\pi^- \pi^+$ are measured using proton--proton collisions corresponding to
$3\mathrm{\,fb}^{-1}$ of integrated luminosity collected at centre-of-mass
energies of $7\mathrm{\,Te\kern -0.1em V}$ and $8\mathrm{\,Te\kern -0.1em
  V}$. The $D^0$ mesons are produced in semileptonic $b$-hadron decays, where
the charge of the accompanying muon is used to determine the initial flavour of
the charm meson. The difference in $C\!P$ asymmetries between the two final
states is measured to be
\begin{align}
  \Delta A_{C\!P} = A_{C\!P}(K^-K^+)-A_{C\!P}(\pi^-\pi^+) = 
  (+0.14 \pm 0.16\mathrm{\,(stat)} \pm 0.08\mathrm{\,(syst)})\% \ . \nonumber
\end{align}
A measurement of $A_{C\!P}(K^-K^+)$ is obtained assuming negligible $C\!P$
violation in charm mixing and in Cabibbo-favoured $D$ decays. It is found to be
\begin{align}
  A_{C\!P}(K^-K^+) = 
  (-0.06 \pm 0.15\mathrm{\,(stat)} \pm 0.10\mathrm{\,(syst)}) \%  \ ,\nonumber 
\end{align}
where the correlation coefficient between $\Delta A_{C\!P}$ and
$A_{C\!P}(K^-K^+)$ is $\rho=0.28$. By combining these results, the $C\!P$
asymmetry in the $D^0\rightarrow\pi^-\pi^+$ channel is
$A_{C\!P}(\pi^-\pi^+)=(-0.20\pm0.19\mathrm{\,(stat)}\pm0.10\mathrm{\,(syst)})\%$.
\end{abstract}
\vspace*{1.0cm}

\begin{center}
  Submitted to JHEP
\end{center}

\vspace{\fill}

{\footnotesize 
\centerline{\copyright~CERN on behalf of the \lhcb collaboration, license \href{http://creativecommons.org/licenses/by/3.0/}{CC-BY-3.0}.}}
\vspace*{2mm}

\end{titlepage}


\newpage
\setcounter{page}{2}
\mbox{~}
\newpage

\centerline{\large\bf LHCb collaboration}
\begin{flushleft}
\small
R.~Aaij$^{41}$, 
B.~Adeva$^{37}$, 
M.~Adinolfi$^{46}$, 
A.~Affolder$^{52}$, 
Z.~Ajaltouni$^{5}$, 
J.~Albrecht$^{9}$, 
F.~Alessio$^{38}$, 
M.~Alexander$^{51}$, 
S.~Ali$^{41}$, 
G.~Alkhazov$^{30}$
% ...

{\footnotesize \it
% ...
$ ^{5}$Clermont Universit\'{e}, Universit\'{e} Blaise Pascal, CNRS/IN2P3, LPC, Clermont-Ferrand, France\\
% ...
$ ^{9}$Fakult\"{a}t Physik, Technische Universit\"{a}t Dortmund, Dortmund, Germany\\
% ...
$ ^{30}$Petersburg Nuclear Physics Institute (PNPI), Gatchina, Russia\\
% ...
$ ^{37}$Universidad de Santiago de Compostela, Santiago de Compostela, Spain\\
$ ^{38}$European Organization for Nuclear Research (CERN), Geneva, Switzerland\\
% ...
$ ^{41}$Nikhef National Institute for Subatomic Physics, Amsterdam, The Netherlands\\
% ...
$ ^{46}$H.H. Wills Physics Laboratory, University of Bristol, Bristol, United Kingdom\\
% ...
$ ^{51}$School of Physics and Astronomy, University of Glasgow, Glasgow, United Kingdom\\
$ ^{52}$Oliver Lodge Laboratory, University of Liverpool, Liverpool, United Kingdom\\
% ...
}
\end{flushleft}

\cleardoublepage


\renewcommand{\thefootnote}{\arabic{footnote}}
\setcounter{footnote}{0}



\pagestyle{plain} 
\setcounter{page}{1}
\pagenumbering{arabic}


%

\section{Introduction}
\label{sec:Introduction}

Decays of charm mesons mediated by the weak interaction provide an attractive
testing ground for physics beyond the Standard Model (SM). Violations of
charge-parity (\CP) symmetry are predicted to be small in charm decays, but
could be enhanced in the presence of non-SM physics. Direct \CP violation arises
when two or more amplitudes with different weak and strong phases contribute to
the same final state. This is possible in singly Cabibbo-suppressed \dkk and
\dpipi decays,\footnote{The inclusion of charge-conjugate processes is implied
  throughout this paper, unless explicitly stated otherwise.} where significant
penguin contributions can be expected~\cite{Bobrowski:2010xg}. Under SU(3)
flavour symmetry, which is approximately valid in heavy quark transitions, the
direct \CP asymmetries in these decays are expected to have equal magnitudes and
opposite sign. For a long time, direct \CP violation in these decays was
expected to be below the $10^{-3}$ level~\cite{Grossman:2006jg}; however, this
prediction has been revisited recently and asymmetries at a few times $10^{-3}$
cannot be excluded within the SM~\cite{Feldmann:2012js, Brod:2011re,
  Bhattacharya:2012ah,LHCb-PAPER-2012-031}. Indirect \CP violation, occurring
through \Dz mixing, is expected to be negligible at the current experimental
precision~\cite{Bianco:2003vb,Grossman:2006jg} and measured to be consistent
with zero~\cite{Amhis:2012bh}. To date, \CP violation in charm decays has not
been established experimentally.

In this paper the \CP asymmetries in \dkk and \dpipi decays are measured in
semileptonic \bquark-hadron decays using the muon charge to identify ({\it tag})
the flavour of the \Dz meson at production. This time-integrated \CP asymmetry
receives contributions from both direct and indirect \CP violation. The
difference of these asymmetries (\DACP) was measured at
LHCb~\cite{LHCb-PAPER-2011-023, *LHCb-CONF-2013-003,
  LHCb-PAPER-2013-003}. Assuming indirect \CP violation to be independent of the
decay mode~\cite{Bianco:2003vb,Grossman:2006jg}, only the effect of direct \CP
violation remains in \DACP. This paper supersedes the previous \DACP result from
Ref.~\cite{LHCb-PAPER-2013-003} that was based on one third of the data. In this
paper the individual \CP asymmetries in \dpipi and \dkk decays are also
measured, using samples of Cabibbo-favoured \Dz and \Dp decays to correct for
spurious asymmetries due to detection and production effects. Individual \CP
asymmetries and their difference have been measured by several other
experiments~\cite{Aubert:2007if, Aaltonen:2011se, *Collaboration:2012qw,
  Staric:2008rx, *Ko:2012px}, all using \Dz mesons tagged by the charge of the
pion from $\Dstarp\to\Dz\pip$ decays. The world average
values~\cite{Amhis:2012bh} are $\AcpKK=(-0.15\pm0.14)\%$ and
$\Acppipi=(0.18\pm0.15)\%$ for the individual asymmetries and
$\DACP=(-0.33\pm0.12)\%$ for the difference.

\section{Method and formalism}
\label{sec:method}

Our procedure to measure the difference in \CP asymmetries, \DACP, follows
Ref.~\cite{LHCb-PAPER-2013-003}. The observed (raw) asymmetry for a \D meson
decay rate to a final state $f$ is defined as
\begin{equation}
  \Araw \equiv \frac{N(\D\to f)-N(\Db\to\fbar)}{N(\D\to f)+N(\Db\to\fbar)} \ ,
\end{equation}
where $N$ is the number of observed decays, \D is either a \Dp or \Dz meson, and
\Db is either a \Dm or \Dzb meson. For decays to a \CP eigenstate, where
$f=\fbar$, the initial flavour of the \Dz meson is tagged by the charge of the
accompanying muon in the semileptonic decay $\Bbar\to\Dz\mun\neumb\PX$, where
\PX denotes possible other particles produced in the decay. Neglecting
third-order terms in the asymmetries, the raw asymmetry in the decays \dkk and
\dpipi is
\begin{equation}
  \Araw = \ACP + \AD + \AP \ ,
  \label{eq:Araw}
\end{equation}
where \AD is any charge-dependent asymmetry in muon reconstruction efficiency
and \AP is the asymmetry between the numbers of \bquark and
\bquarkbar hadrons (denoted as \Bbar and \B, respectively) produced in the LHCb
acceptance, which includes possible \CP violation in \Bd mixing. The production
and detection asymmetries are common to the \dkk and \dpipi decay modes, so they
cancel in the difference of the raw asymmetries, giving
\begin{equation}
  \DACP =  \AKK - \Apipi = \AcpKK - \Acppipi \ .
  \label{eq:delta}
\end{equation}

The production and muon detection asymmetry in Eq.~(\ref{eq:Araw}) can also be
removed using the Cabibbo-favoured \dkpi decay mode in
$\Bbar\to\Dz\mun\neumb\PX$ decays. In this decay, \CP violation can be neglected
as it is expected to be significantly suppressed compared to our sensitivity for
measuring \CP violation in \dkk and \dpipi decays. In the reconstruction of the
$\Km\pip$ final state there is an instrumental asymmetry, \ADKpi, due to the
different interaction cross section of positively and negatively charged kaons
in the detector material. Also, other detector-related effects, for example due
to the acceptance, selection and detection inefficiencies, can contribute to
this detection asymmetry. The raw asymmetry in this decay mode is then
\begin{equation}
  \AKpi = \AD + \AP +\ADKpi \ .
\label{eq:Arawkpi}
\end{equation}
The detection asymmetry \ADKpi of the final state $\Km\pip$ is obtained from \Dp
decays produced directly in $pp$ collisions (so-called {\it prompt} \Dp
decays). Two decay modes are used, \dkpipi and \dkzpi with
$\Kzb\to\pip\pim$. The raw asymmetry of \dkpipi decays is
\begin{equation}
  \ArawKpipi = \APD + \ADKpi + \ADpi \ ,
\label{eq:Arawkpipi}
\end{equation}
where \APD is the production asymmetry of prompt \Dp mesons and \ADpi is the
detection asymmetry of the other charged pion. The raw asymmetry of \dkzpi
decays is
\begin{equation}
  \ArawKzpi = \APD + \ADpi - \ADkz \ ,
\label{eq:Arawkspi}
\end{equation}
where \ADkz is the detection asymmetry of the decay \kzpipi, which is discussed
later. Taking the difference between Eqs.~(\ref{eq:Arawkpipi}) and
(\ref{eq:Arawkspi}), \ADKpi is obtained as
\begin{equation}
  \ADKpi = \ArawKpipi - \ArawKzpi - \ADkz \ .
\label{eq:ADKpi}
\end{equation}
This method assumes negligible \CP violation in these Cabibbo-favoured \Dp decay
modes. By combining Eqs.~(\ref{eq:Araw}) and (\ref{eq:Arawkpi}), the \CP
asymmetry in the \dkk decay becomes
\begin{equation}
  \AcpKK = \ArawKK - \AKpi + \ADKpi \ ,
\label{eq:Acp}
\end{equation}
where \ADKpi is taken from Eq.~(\ref{eq:ADKpi}). The \CP asymmetry in the \dpipi
decay is determined from the difference between the measurements of \AcpKK and
\DACP.

\section{Detector}
\label{sec:Detector}

The \lhcb detector~\cite{Alves:2008zz} is a single-arm forward spectrometer
covering the \mbox{pseudorapidity} range $2<\eta <5$, designed for the study of
particles containing \bquark or \cquark quarks. The detector includes a
high-precision tracking system consisting of a silicon-strip vertex detector
surrounding the $pp$ interaction region, a large-area silicon-strip detector
located upstream of a dipole magnet with a bending power of about $4{\rm\,Tm}$,
and three stations of silicon-strip detectors and straw drift tubes placed
downstream of the magnet. The polarity of the magnetic field is regularly
reversed during data taking.  The combined tracking system provides a momentum
measurement with relative uncertainty that varies from 0.4\% at low momentum,
$p$, to 0.6\% at 100\gevc, and impact parameter resolution of 20\mum for charged
particles with large transverse momentum, \pt. Different types of charged
hadrons are distinguished by information from two ring-imaging Cherenkov
detectors~\cite{LHCb-DP-2012-003}. Photon, electron and hadron candidates are
identified by a calorimeter system consisting of scintillating-pad and preshower
detectors, an electromagnetic calorimeter and a hadronic calorimeter. Muons are
identified by a system composed of alternating layers of iron and multiwire
proportional chambers. The trigger~\cite{LHCb-DP-2012-004} consists of a
hardware stage, based on information from the calorimeter and muon systems,
followed by a two-stage software stage, that applies a full event
reconstruction.

\section{Data set and selection}
\label{sec:selection}

This analysis uses the data set collected by \lhcb corresponding to an
integrated luminosity of $3\invfb$. The data in 2011 ($1\invfb$) were taken at a
centre-of-mass energy of 7\tev and the data in 2012 ($2\invfb$) were taken at a
centre-of-mass energy of 8\tev. The fraction of data collected with up (down)
polarity of the magnetic field is $40\%$ ($60\%$) in 2011 and $52\%$ ($48\%$) in
2012.  Charge-dependent detection asymmetries originating from any left-right
asymmetry in the detector change sign when the field polarity is reversed. By
design, the analysis method does not rely on any cancellation due to the regular
field reversals, since all detection asymmetries are already removed in the
determination of the \CP asymmetries.  This assumption is tested by performing
the analysis separately for the two polarities. To ensure that any residual
detection asymmetries cancel, the raw asymmetries are determined from the
arithmetic mean of the results obtained for the two magnet
polarities. Similarly, the analysis is performed separately for the 2011 and
2012 data as detection asymmetries and production asymmetries change due to
different operational conditions.

At the hardware trigger stage, the events in the semileptonic \B decay modes are
required to be triggered by the muon system. The muon transverse momentum must
be larger than $1.64\gevc$ for the 2011 data and larger than $1.76\gevc$ for the
2012 data.  In the software trigger, the muon candidate is first required to
have $\pt>1.0\gevc$ and a large impact parameter. Then, the muon and one or two
of the \Dz decay products are required to be consistent with the topological
signature of \bquark-hadron decays~\cite{LHCb-DP-2012-004}. Since the \DACP
measurement has no detection asymmetry coming from the \Dz decay products, \B
candidates triggered on the presence of a final-state particle with high \pt and
large impact parameter are also accepted in the corresponding selection.

The remaining selection of semileptonic \B decays reduces the background from
prompt \Dz decays to the per-cent level. The residual background consists mainly
of combinations from inclusive \bquark-hadron decays with other particles in the
event. The selection is similar to that in the previous
publication~\cite{LHCb-PAPER-2013-003}, except for the looser particle
identification requirements of the kaon candidates. To reduce the large \dkpi
sample size only half of the candidates (randomly selected) is kept.

In order to reduce possible biases induced by trigger criteria, the events in
the prompt charm decay modes are selected by the hardware trigger, independently
of the presence of the \Dp candidate. In the software trigger, one of the
final-state pions is first required to have $\pt>1.6\gevc$ and a large
impact parameter to any primary vertex. This ensures that the distributions of
the kaon and the other pion in the \dkpipi decay and of the \Kzb meson in the
\dkzpi decay are not biased by these trigger requirements. Finally, an exclusive
selection is applied for each \Dp decay mode in the last stage of the software
trigger. This is similar to the offline selection, where a secondary vertex is
reconstructed and required to be significantly displaced from any primary
vertex.

All particles are required to have $p>2\gevc$ and $\pt>250\mevc$. Additionally,
all tracks in the \dkpipi and \dkzpi decays are required to have a large impact
parameter with respect to any primary vertex.  The particle identification
requirements are the same as those in the \dkpi decay mode from semileptonic \B
decays. The neutral kaon in the \dkzpi decay is detected in the $\pip\pim$ final
state, which is dominated by the decay of the \KS state. When \KS mesons decay
early such that both pions leave sufficiently many hits in the vertex detector
and in the three downstream tracking stations, the pions can be reconstructed as
so-called {\em long} tracks. When \KS mesons decay later such that both pions do
not leave enough hits in the vertex detector, but enough hits in the rest of the
tracking system, the pions can be reconstructed as so-called {\em downstream}
tracks. Downstream \KS candidates are available only in 2012 data, since no
dedicated trigger was available to select these decays in 2011. For this reason,
only \dkzpi decays formed with long \KS candidates are used for the asymmetry
measurement. The downstream \KS candidates are used to check the effect of the
\Kz detection asymmetry. There is no \pt requirement for the pions from
downstream \KS candidates. All \KS candidates are required to have a large
impact parameter. Both \Dp and \KS candidates are required to have $\pt>1\gevc$
and an accurately reconstructed decay vertex. For the \dkpipi decay the scalar
\pt sum of the \Dp daughters is required to be larger than $2.8\gevc$. The \Dp
candidates should have a large impact parameter and significant flight distance
from the primary vertex. Given the large branching fraction of the \dkpipi
decay, only one fifth of the available data set (randomly selected) is
considered in the following.

To improve the mass resolution, a vertex fit~\cite{Hulsbergen:2005pu} of the \Dp
decay products is made, where the \Dp candidate is constrained to originate from
the corresponding primary vertex. Additionally, in the decay \dkzpi the mass of
the \Kzb meson is constrained to the nominal value~\cite{PDG2012}. The momentum
of all particles is corrected~\cite{LHCB-PAPER-2013-011} to improve the
stability of the mass scale versus data taking period and to reduce the width of
the mass distribution.

\section{Determination of the asymmetries}
\label{sec:results}

In this section the raw asymmetries are obtained from fits to the invariant mass
distributions. These numbers are corrected for effects coming from the \Kz
detection asymmetry and wrong flavour tags. The contributions from direct and
indirect \CP violation are determined and finally the \CP asymmetries are
calculated.

\subsection{Invariant mass distributions}
\label{sec:fitmodel}

\begin{figure}
  \begin{center}
    \includegraphics[width=0.49\textwidth]{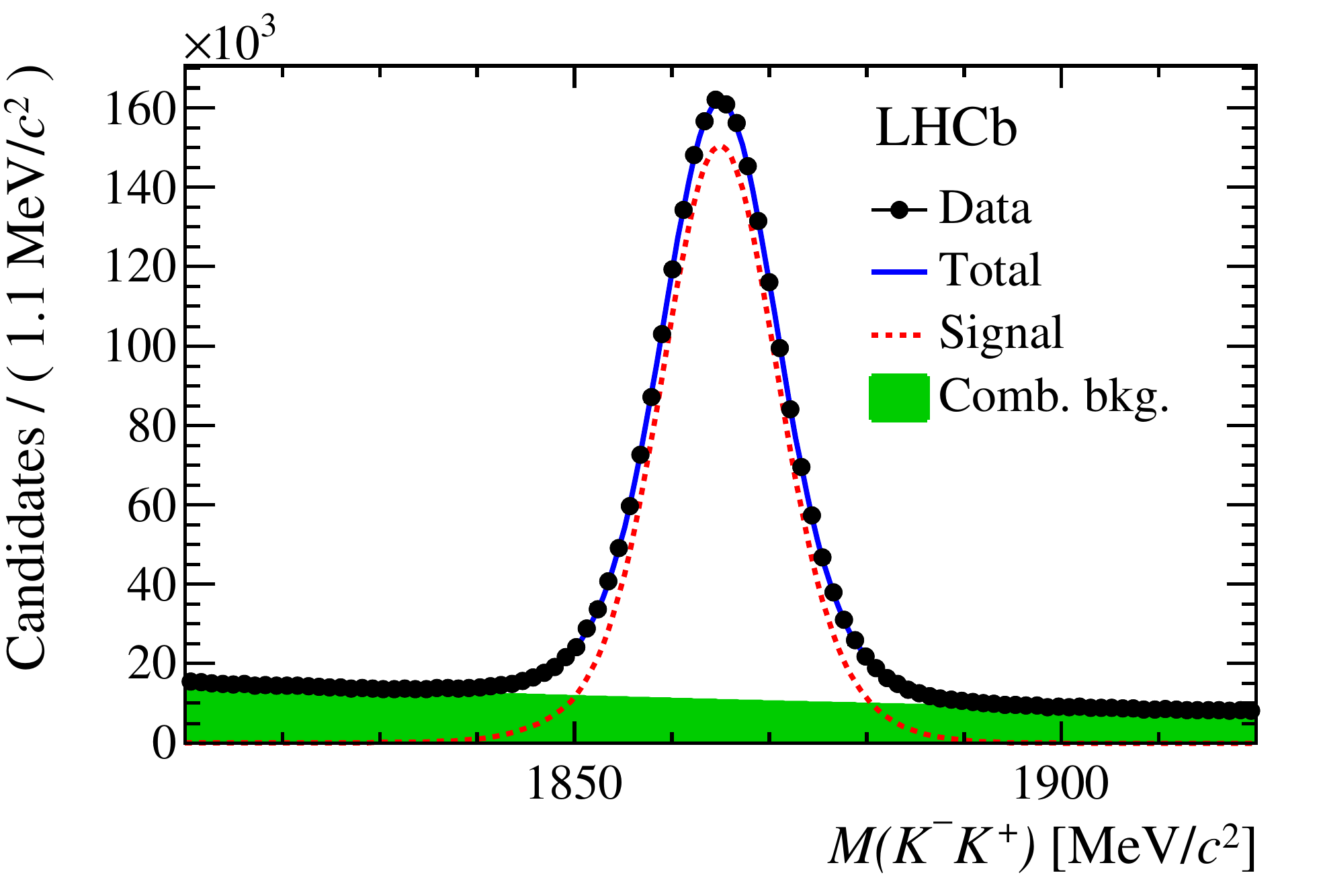}\put(-170,120){(a)}
    \includegraphics[width=0.49\textwidth]{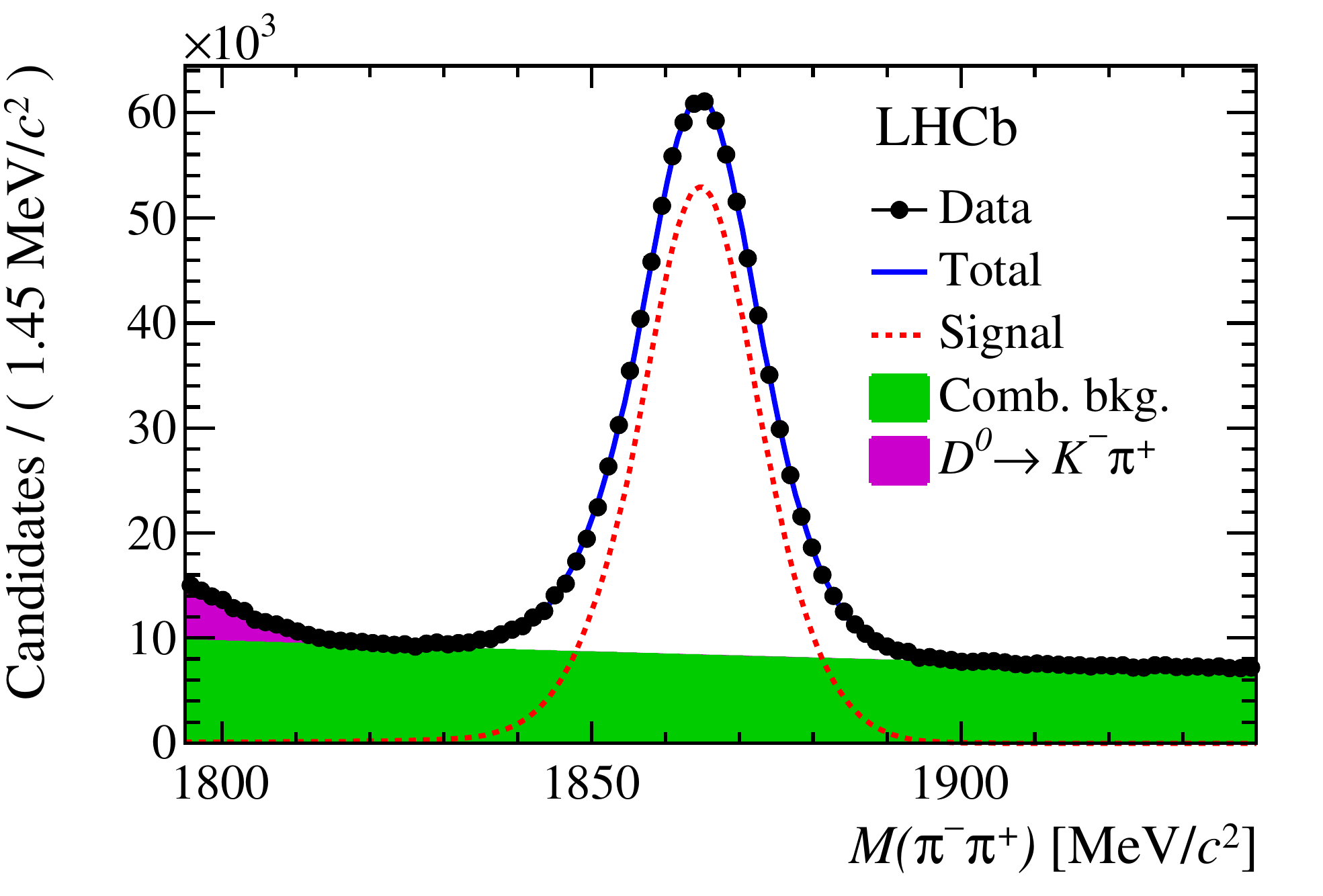}\put(-170,120){(b)}
  \end{center}
  \begin{center}
    \includegraphics[width=0.49\textwidth]{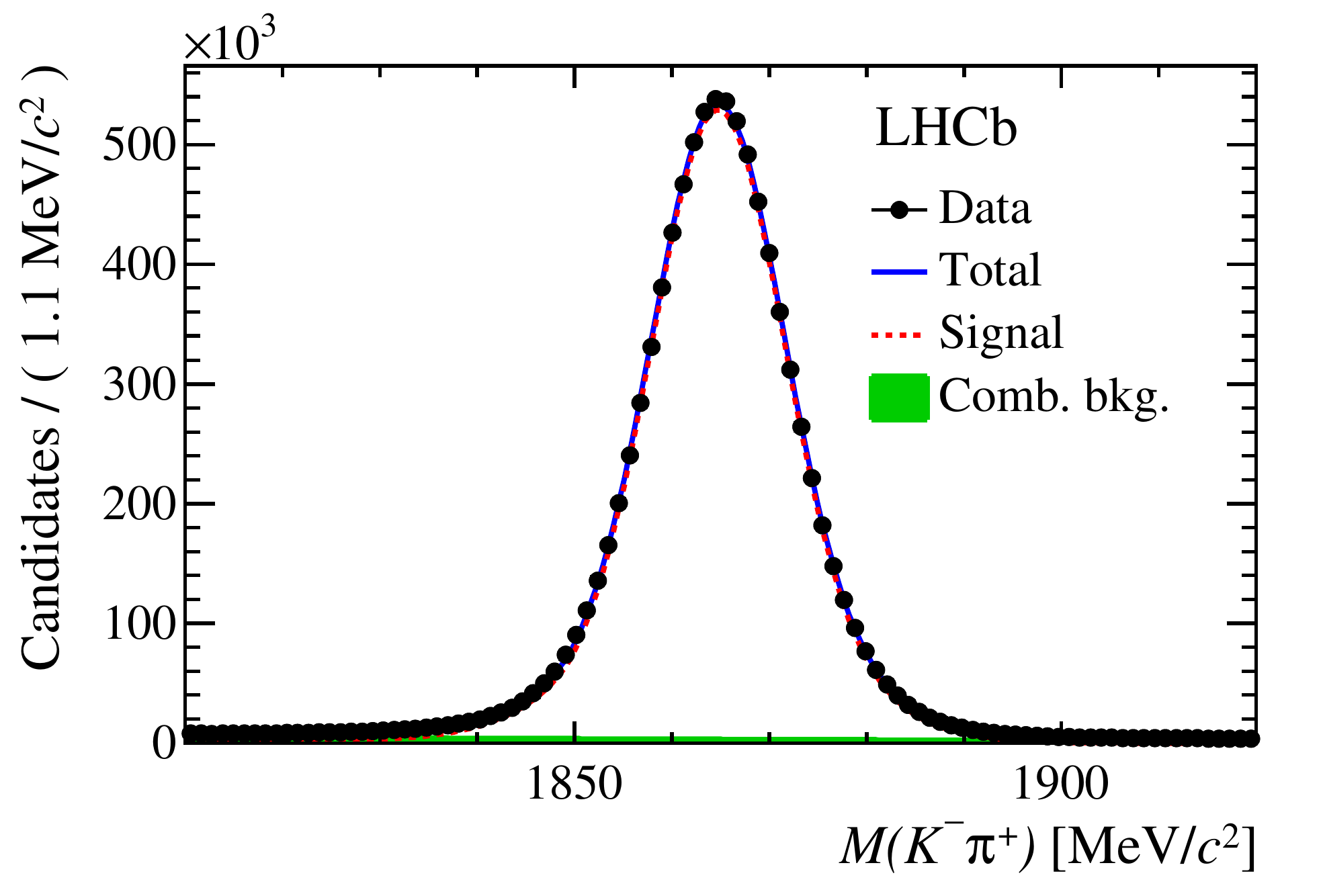}\put(-170,120){(c)}
    \includegraphics[width=0.49\textwidth]{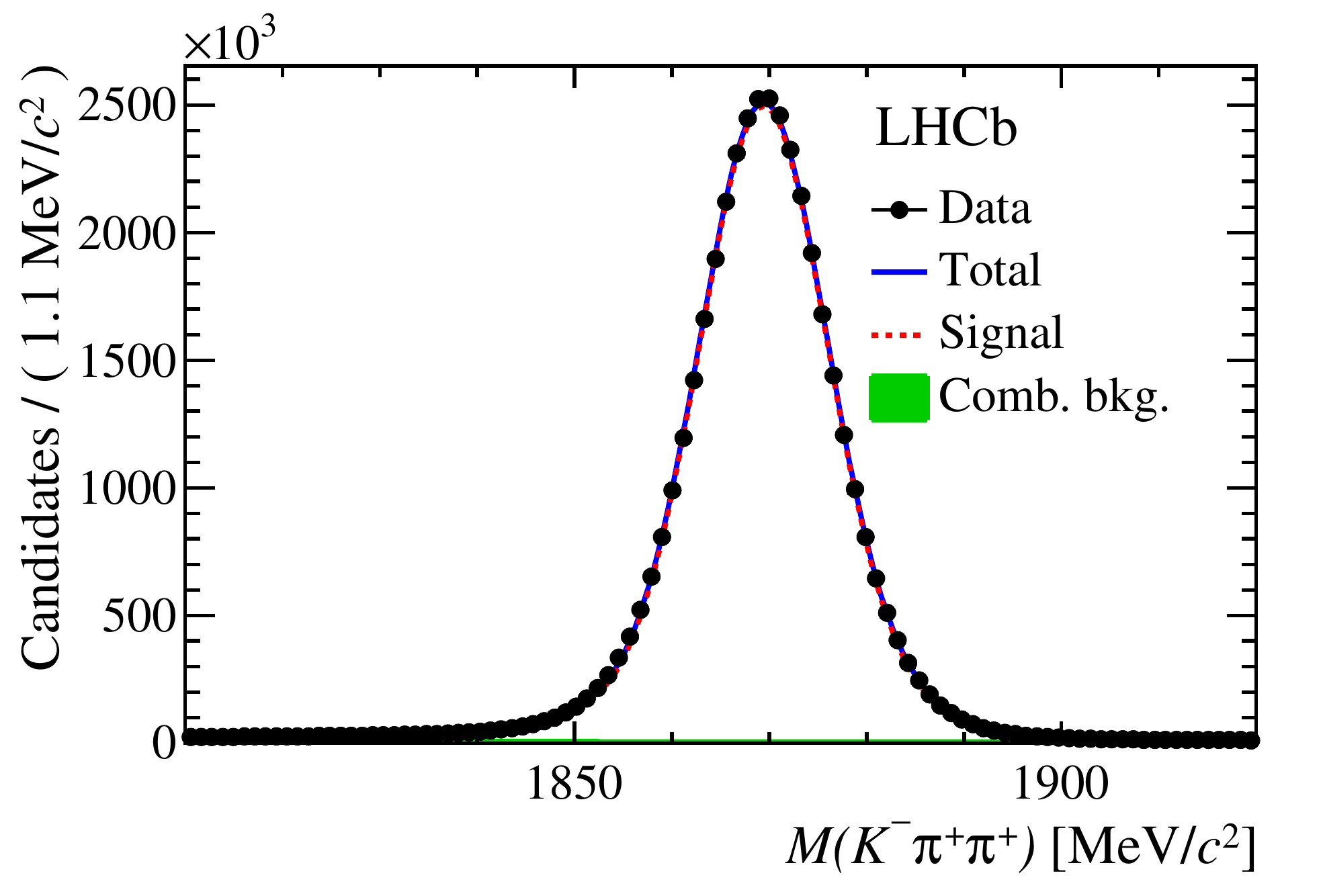}\put(-170,120){(d)}
  \end{center}
  \begin{center}
    \includegraphics[width=0.49\textwidth]{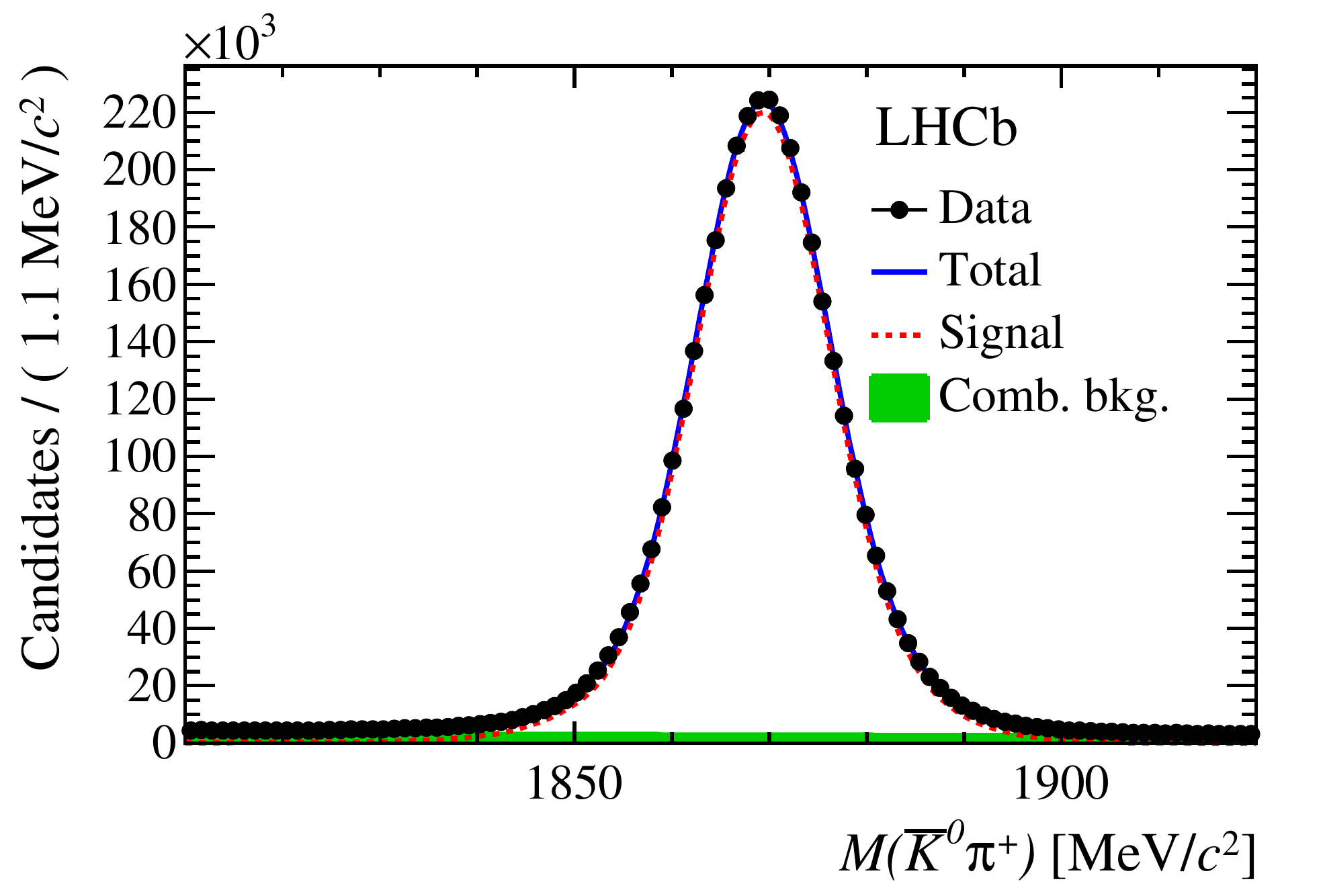}\put(-170,120){(e)}
  \end{center}
  \vspace*{-0.5cm}
  \caption{Invariant mass distributions for muon-tagged (a) \dkk, (b) \dpipi and
    (c) \dkpi candidates and for prompt (d) \dkpipi and (e) \dkzpi candidates.
    The results of the fits are overlaid.}
  \label{fig:plot_mass}
\end{figure}

Invariant mass distributions for the \Dz and \Dp candidates are shown in
Fig.~\ref{fig:plot_mass} with the fit results overlaid. For all decay modes the
signal is modelled by the sum of two Gaussian functions with common mean
and a power-law tail. For decays to non-\CP eigenstates (i.e., \dkpi, \dkpipi,
\dkzpi) different means and average widths are allowed between \D and \Db
states, due to a known charge-dependent bias in the measurement of the
momentum. The background is described by an exponential function, with different
slopes for \D and \Db states. An overall asymmetry in the number of
background events is also included in the model. The background from
misidentified \dkpi decays in the \dpipi invariant mass distribution is modelled
with a single Gaussian function with the same shape for both muon tags and an
additional asymmetry parameter. The numbers of signal decays determined from
fits to the invariant mass distributions are given in Table~\ref{tab:yields}.

\begin{table}
  \begin{center}
    \caption{Number of signal decays determined from fits to the invariant
      mass distributions.}
  \label{tab:yields}
  \vspace{0.1cm}
  \begin{tabular}{l r} \hline
Decay sample & Signal decays \\ 
\hline
\dpipi from \B                 &   773 541    \\ 
\dkk from \B, \DACP selection  &  2 166 045   \\
\dkk from \B, \AcpKK selection &  1 821 462   \\
\dkpi from \B                  &   9 088 675  \\ 
Prompt \dkpipi                 & 40 782 645   \\ 
Prompt \dkzpi, long \KS        &  3 765 530   \\ 
Prompt \dkzpi, downstream \KS  &  2 512 615   \\ \hline
\end{tabular}
\end{center}
\end{table}

\subsection{Differences in kinematic distributions}
\label{sec:weighting}

Production and detection asymmetries depend on the kinematic distributions of
the particles involved. Since the momentum distributions of the particles in the
signal and calibration decay modes are different, small residual production and
detection asymmetries can remain in the calculation of \DACP and \AcpKK. This
effect is mitigated by assigning weights to each candidate such that the
kinematic distributions are equalised. For the measurement of \DACP, the \dkk
candidates are weighted according to the \pt and $\eta$ values of the \Dz, which
are the kinematic variables showing the most significant differences. The
weights are chosen such that the weighted and background-subtracted
distributions of the \Dz and muon candidates agree with the corresponding
(unweighted) distributions in the \dpipi sample. After weighting, the effective
sample size is given by $N_{\rm eff}=(\sum_{i=1}^Nw_i)^2/(\sum_{i=1}^Nw_i^2)$,
where $w_i$ is the weight of candidate $i$ and $N$ the total number of
candidates. Due to the good agreement in the kinematic distributions between the
two decays, this procedure reduces the statistical power of the weighted \dkk
event sample by only $8\%$.

For the measurement of \AcpKK, additional weighting steps for the \Dp
calibration decay modes are needed. In the first step, the \dkpi candidates are
weighted based on the \pt and $\eta$ values of the \Dz meson, such that they
agree with the corresponding (unweighted) distributions of the \dkk
candidates. This step, which reduces the statistical power of the \dkpi sample
by $3\%$, ensures the cancellation of the \B production asymmetry and muon
detection asymmetry. In the second step, the \dkpipi candidates are weighted
according to the \pt and $\eta$ values of both the kaon and the pion that was
not selected by the software trigger. This step equalises the kinematic
distributions of the \Km and \pip to those of the (now weighted) \dkpi decay to
ensure cancellation of the $\Km\pip$ detection asymmetry. The resulting $50\%$
reduction in statistical power does not contribute to the final uncertainty
given the large number of \dkpipi candidates available.  In the last step, the
\dkzpi candidates are weighted according to the \pt and $\eta$ values of both
the pion and the \Dp candidate, such that they agree with the corresponding
(weighted) distributions of the \dkpipi candidates. The last step ensures
cancellation of the \Dp production asymmetry and detection asymmetry from the
pion that is used in the software trigger and reduces the statistical power of
the \dkzpi sample by $77\%$.

\subsection[K0 asymmetry]{{\boldmath \Kz} asymmetry}
\label{sec:KzAsymmetry}

An asymmetry in the detection of a \Kz to the $\pip\pim$ final state arises from
the combined effect of \CP violation and mixing in the neutral kaon system and
the different interaction rates of \Kz and \Kzb in the detector material. Due to
material interactions, a pure \KL state can change back into a superposition of
\KL and \KS states~\cite{Pais:1955sm}. These regeneration and \CP-violating
effects are of the same order and same sign in LHCb. To estimate the total \Kz
detection asymmetry, the mixing, \CP violation and absorption in material
need to be described coherently. The amplitudes in the \KL and \KS basis of an
arbitrary neutral kaon state in matter evolve as~\cite{Fetscher:1996fa}
\begin{align}
  \alphaL(t) &= \textrm{e}^{-i \, t \, \Sigma}
  \left[ \alphaL(0) \cos{(\Omega t )} - i\frac{ \alphaL(0) \Delta \lambda +
      \alphaS(0) \Delta \chi}{2 \Omega} \sin{(\Omega t )} \right] \ ,\\
  \alphaS(t) &= \textrm{e}^{-i \, t \, \Sigma}
  \left[ \alphaS(0) \cos{(\Omega t )} + i\frac{ \alphaS(0) \Delta \lambda -
  \alphaL(0) \Delta \chi}{2 \Omega} \sin{(\Omega t )} \right] \ ,
\label{eq:truf}
\end{align}
where the constants $\Omega \equiv \frac{1}{2} \sqrt{\Delta \lambda^{2}+\Delta
  \chi^{2}}$ and $\Sigma \equiv \frac{1}{2}(\lambdaL + \lambdaS + \chi +
\bar{\chi})$ are given by the masses $\mLS$ and decay widths $\GammaLS$
of the \KL and \KS states and by the absorption $\chi$ ($\bar{\chi}$) of \Kz
(\Kzb) states through
\begin{align}
\Delta \lambda &= \lambdaL - \lambdaS 
                = \Delta m - \frac{i}{2} \Delta \Gamma=
                    (\mL-\mS) - \frac{i}{2}(\GammaL - \GammaS) \ ,\nonumber \\
\Delta \chi    &= \chi - \bar{\chi} = -\frac{2\pi \mathcal{N}}{m}(f-\bar{f}) 
                = -\frac{2\pi \mathcal{N}}{m}{\Delta f} \ ,
\label{eq:good}
\end{align}
where $\mathcal{N}$ is the scattering density, $m$ the kaon mass, and $f$ and
$\bar{f}$ the forward scattering amplitudes. The imaginary part of $f$ is
related to the total cross section through the optical theorem $\sigma_T =
(4\pi/p) {\rm Im} f$. The difference in the interaction cross sections of \Kzb
and \Kz depends on the momentum of the kaon and on the number of nucleons,
$A$, in the target and is obtained from Ref.~\cite{Gsponer:1978dt},
\begin{equation}
\Delta \sigma = \sigma_T(\Kzb) - \sigma_T(\Kz) 
              = 23.2 \, A^{0.758}[p(\gevcNS)]^{-0.614} \mbarn \ .
\label{eq:Gsponer}
\end{equation}
The phase of $\Delta f$ is determined using the phase-power
relation~\cite{Gsponer:1978dt, Briere:1995tw} to be $\arg(\Delta
f)=(-124.7\pm0.8)\degrees$. The regeneration incorporates two
effects~\cite{Pais:1955sm, Good:1957zza}. The term ${\rm Im}(\Delta f)$
describes the incoherent regeneration due to absorption and elastic scattering,
which is equivalent to the case of charged kaons. The term ${\rm Re}(\Delta f)$
describes the coherent regeneration due to dispersion (phase shift) of the \Kz
and \Kzb states.

As the neutral kaons are produced in a flavour eigenstate, the initial
amplitudes at $t=0$, $\alphaLS(0)$, need to be written in the \KL and
\KS basis,
\begin{equation}
  \ket{\Kz}, \ket{\Kzb} = \sqrt{\frac{1+|\epsilon|^2}{2}} 
  \frac{1}{1\pm\epsilon} \left[\ket{\KL} \pm \ket{\KS} \right] \ ,
\label{eq:kzkzb} 
\end{equation}
where $\epsilon$ describes \CP violation in kaon mixing. At a given time, the
decay rate into the final state $\pip\pim$ is given by $|\alphaS(t) +
\epsilon\, \alphaL(t)|^2$. The values of the parameters used to calculate
the \Kz asymmetry are given in Table~\ref{tab:k0Params}.

\renewcommand{\arraystretch}{1.1}
\begin{table} 
  \begin{center} 
    \caption{Values of the parameters used to calculate the \Kz
      asymmetry~\cite{PDG2012,Gsponer:1978dt}.}
    \label{tab:k0Params}
    \vspace{0.1cm}
    \begin{tabular}{l c}\hline 
 Parameter              & Value \\ \hline 
 $\Delta m$             & $(0.5293 \pm 0.0009)\times10^{10}\hbar\sec^{-1}$\\
 $\tauS\equiv1/\GammaS$ & $(0.8954\pm0.0004)\times10^{-10}\sec$\\
 $\tauL\equiv1/\GammaL$ & $(5.116\pm0.021 )\times10^{-8} \sec $\\
 $m$                    & $ (497.614\pm0.024)\mevcc $  \\
 $\arg(\Delta f)$       & $(-124.7\pm 0.8)^{\circ}$ \\
 $|\epsilon|$           & $(2.228 \pm 0.011)\times10^{-3}$ \\
 $\phi_{+-}\equiv\arg\epsilon$ & $(43.51\pm0.05)^{\circ} $ \\
      \hline
    \end{tabular}
  \end{center}
\end{table}
\renewcommand{\arraystretch}{1.0}

Using the \KS and \Dz decay positions, the path of the \KS meson through the
detector is known and the expected \Kz asymmetry can be calculated using the
formulae above. For every \dkzpi candidate, the path of the neutral kaon is
divided into small steps using the material model of the LHCb detector. At each
step, the amplitudes are updated using Eq.~(\ref{eq:truf}) starting with either
a \Kz or \Kzb as initial state. The expected \Kz asymmetry for a given event is
then the asymmetry in the decay rates between the \Kz and \Kzb initial
states. The overall asymmetry is calculated from the expected asymmetry averaged
over all reconstructed \dkzpi candidates.

\begin{figure}
\begin{center}
\includegraphics[width=0.49\textwidth]{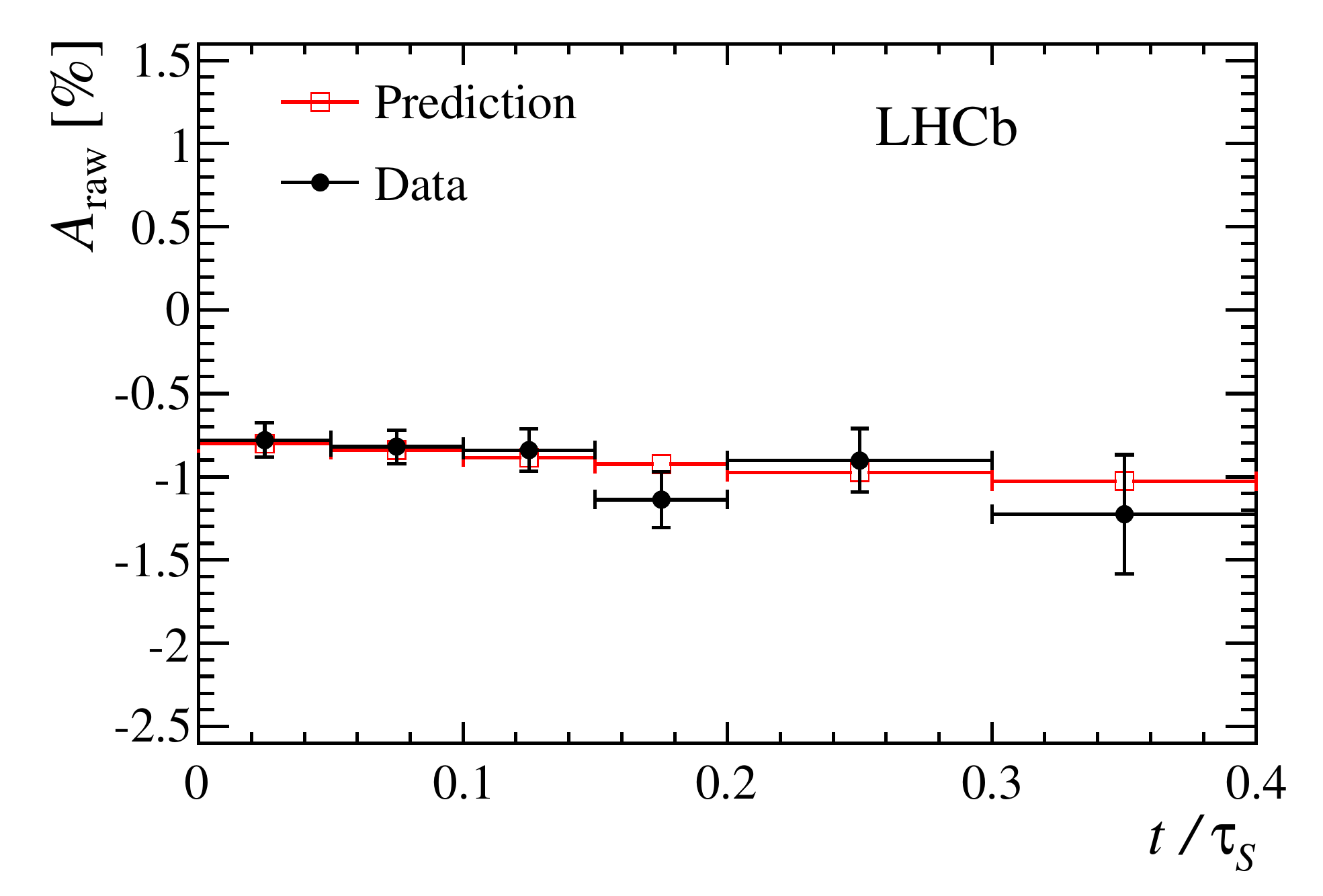}
\put(-35,125){(a)} 
\put(-75,105){\small Long \KS} 
\includegraphics[width=0.49\textwidth]{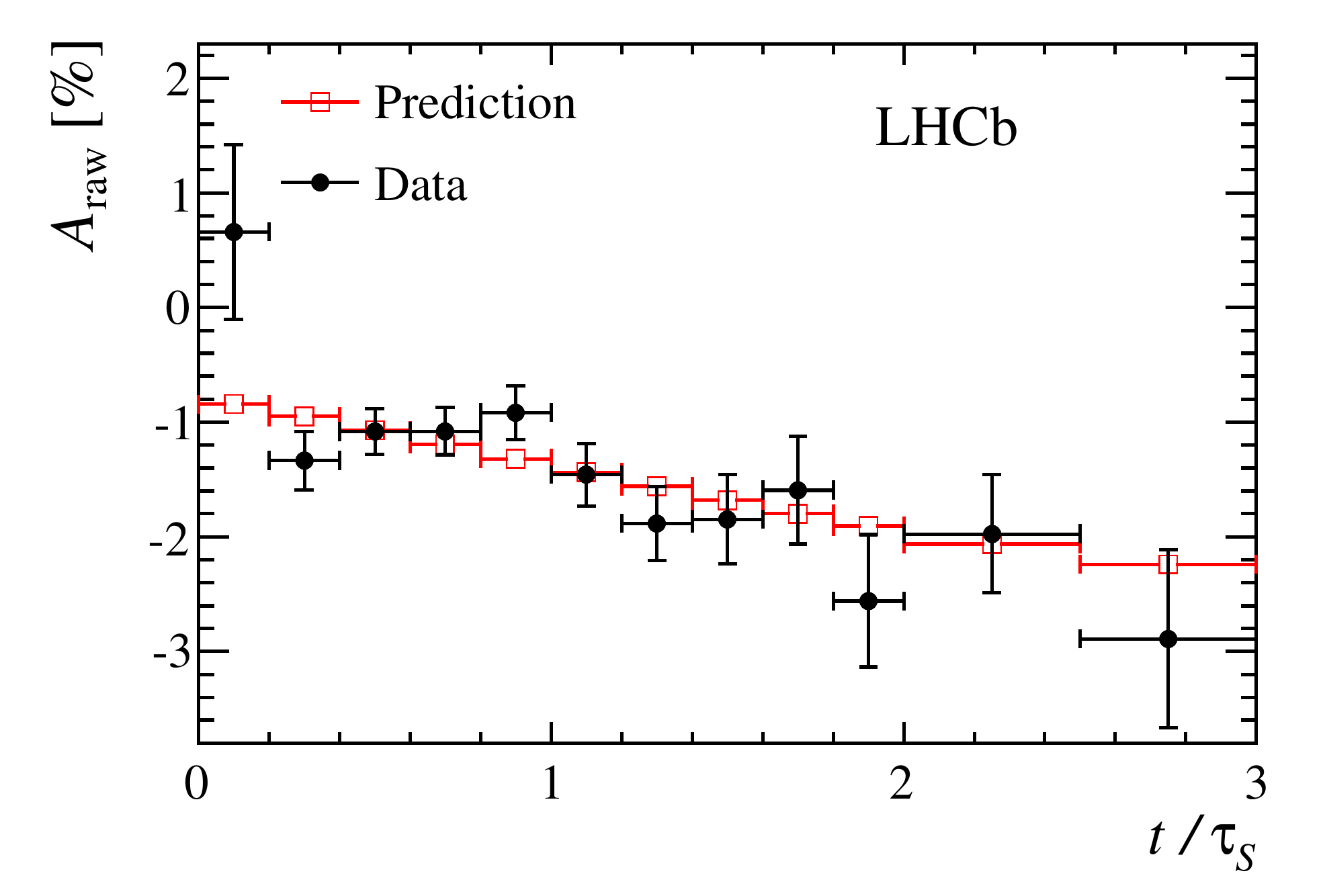} 
\put(-35,125){(b)}
\put(-110,105){\small Downstream \KS}
 \caption{\small Raw asymmetry in the \dkzpi decay shown for (a) long and (b)
   downstream \KS candidates versus the \KS decay time in units of its
   lifetime. The long \KS candidates are reconstructed in the full data set,
   while the downstream \KS candidates are reconstructed in the 2012 data
   only. The predicted effect from the \Kz asymmetry, $-\ADkz$, is also shown.
   An overall shift is applied to this prediction to account for \Dp production
   and pion detection asymmetries (note that the unshifted \ADkz at $t=0$ is
   zero).}
 \label{fig:comparisonKs} 
\end{center} 
\end{figure}

The measured raw asymmetry in the \dkzpi decay and the effect from the predicted
\Kz asymmetry are shown as functions of the \KS decay time in
Fig.~\ref{fig:comparisonKs}. The measured raw asymmetry receives contributions
not only from the \Kz detection asymmetry, but also from the pion tracking
asymmetry and \Dp production asymmetry. These contributions are almost
independent of the \KS decay time. Therefore, an overall shift is applied to the
predicted asymmetry to match the data. Assuming a negligible pion detection
asymmetry, this shift agrees well with the \Dp production asymmetry of
$(-0.96\pm0.26)\%$ measured on 2011 data~\cite{LHCb-PAPER-2012-026}. The
downward trend coming from the \Kz asymmetry is clearly visible, in particular
for downstream \KS decays. The predicted asymmetry dependence agrees well
with the data, with $p$-values of 0.81 and 0.31, respectively.

Only \KS candidates reconstructed with long tracks are used in the \AcpKK
measurement. These candidates probe lower \KS decay times, compared to those
reconstructed with downstream tracks, resulting in a much smaller \Kz asymmetry
correction. Nevertheless, the effect observed in downstream \KS decays is used
to test the accuracy of the \Kz detection asymmetry model. The measured
difference in raw asymmetry between the samples is $(0.49 \pm 0.12)\%$, which is
obtained after weighting the long \KS sample to correct for differences in the
\Dp production and pion detection asymmetries. This value agrees with the
expected difference of $(0.546 \pm 0.027)\%$, where the uncertainty is dominated
by the uncertainty on the amount of detector material. The relative uncertainty
of the measured difference ($25\%$) is assigned as a systematic
uncertainty on the \Kz asymmetry model. The expected \Kz asymmetry in the \dkzpi
sample with long \KS candidates weighted according to the procedure described in
Sect.~\ref{sec:weighting} is found to be $\ADkz=(0.054 \pm 0.014\syst)\%$.

\subsection{Wrong flavour tags}
\label{sec:mistag}

If a \Dz meson is combined with a muon that does not originate from the
corresponding semileptonic \B decay, the \Dz flavour may not be correctly
assigned. The probability to wrongly tag a \Dz meson is denoted by \mistag. This
mistag probability dilutes the observed asymmetry by a factor $1-2\mistag$. For
small \mistag, the expression of \DACP can be written as
\begin{equation}
  \DACP =  (1+2\mistag)[\AKK - \Apipi] \ .
  \label{eq:mistagDACP}
\end{equation}
The mistag probability only affects the semileptonic decay modes as the flavour
of the \Dp reconstruction is unambiguous.  In the reconstruction of the \dkpi
decay, wrong-sign decays coming from doubly Cabibbo-suppressed $\Dz\to\Kp\pim$
decays and mixed $\Dz\to\decay{\Dzb}{\Kp\pim}$ decays are included. Hence, the
calculation of the \CP asymmetry of the \dkk decay is
\begin{equation}
  \AcpKK = (1+2\mistag)[\ArawKK - \AKpi] + (1-2R)\ADKpi \ ,
  \label{eq:mistagKK}
\end{equation}
where $R$ is the ratio of branching fractions of wrong-sign $\Dz\to\Kp\pim$
decays over right-sign \dkpi decays.

The \dkpi sample from semileptonic \B decays is also used to estimate the mistag
probability. The final state, either $\Kp\pim$ or $\Km\pip$, almost
unambiguously determines the flavour of the \Dz meson, since the contamination
from wrong-sign decays is only
$R=(0.389\pm0.003)\%$~\cite{LHCb-PAPER-2013-053}. After correcting for this
wrong-sign fraction, the mistag probability is found to be
$\mistag=(0.988\pm0.006)\%$ for the \DACP measurement and
$\mistag=(0.791\pm0.006)\%$ for the \AcpKK measurement. The small difference
between these numbers is due to more stringent trigger criteria in the latter,
resulting in different kinematic distributions for the two selections. As a
consistency check, the mistag probability is obtained in all three semileptonic
samples by searching for an additional pion from a $\Dstarp\to\Dz\pip$ decay and
comparing the charge of this pion with that of the muon. The mistag
probabilities are found to be in good agreement with an average value of
$\mistag=(0.985\pm0.017)\%$ for the \DACP measurement and of
$\mistag=(0.803\pm0.019)\%$ for the \AcpKK measurement. Since the dilution
effect from such a small \mistag value results in tiny corrections to \DACP and
\AcpKK, the uncertainty in this number is neglected. A small difference of
$\Delta\mistag=(0.028\pm0.011)\%$ is observed between the probabilities to
wrongly tag \Dz and \Dzb mesons. Although any non-zero value is expected to
cancel in the calculation of \DACP and \AcpKK, the full difference is
conservatively taken as a systematic uncertainty.

\subsection{Average decay times}

The time-integrated \CP asymmetry has contributions from direct and indirect \CP
violation, depending on the average decay time, $\mean{t}$, of the \Dz mesons in
the sample as~\cite{Grossman:2006jg}
\begin{equation}
  A_{\CP} \approx a_{\CP}^{\rm dir} - A_{\Gamma} \frac{\mean{t}}{\tau} \ ,
 \label{eq:AcpDirectIndirect}
\end{equation}
where $a_{\CP}^{\rm dir}$ is the direct \CP violation term, $\tau$ the \Dz
lifetime, and $A_{\Gamma}$ a measure of indirect \CP violation. The
world-average value of $A_{\Gamma}$ in singly Cabibbo-suppressed \Dz decays is
$(-0.014\pm0.052)\%$~\cite{Amhis:2012bh}. Assuming that this quantity is the
same for \dkk and \dpipi decays, the sensitivity of \DACP to indirect \CP
violation is introduced by the difference in the average \Dz decay times between
the two decay modes. A complete discussion is given in the previous
publication~\cite{LHCb-PAPER-2013-003} and the same procedure is adopted
here. In this procedure, the average decay time of the signal in each
sample is determined by subtracting the decay time distributions of background
events using the \sPlot\ technique~\cite{Pivk:2004ty}, and by correcting for
decay time resolution effects. For the \DACP measurement, the average decay
times are found to be
\begin{align}
  \mean{t}/\tau(\Km\Kp)   &= 1.082\pm0.001\stat\pm0.004\syst \ , \nonumber\\ 
  \mean{t}/\tau(\pim\pip) &= 1.068\pm0.001\stat\pm0.004\syst \ . \nonumber 
\end{align}
The small difference between these numbers ($0.014\pm0.004$) implies that
$\DACP=\Delta a_{\CP}^{\rm dir}$ is an excellent approximation. For the \AcpKK
measurement, the average decay time is found to be
\begin{equation}
  \mean{t}/\tau(\Km\Kp) = 1.051\pm0.001\stat\pm0.004\syst \ . \nonumber
\end{equation}

\subsection[CP asymmetry measurements]{{\boldmath \CP} asymmetry measurements}

\begin{figure}
  \begin{center}
    \includegraphics[width=0.78\textwidth]{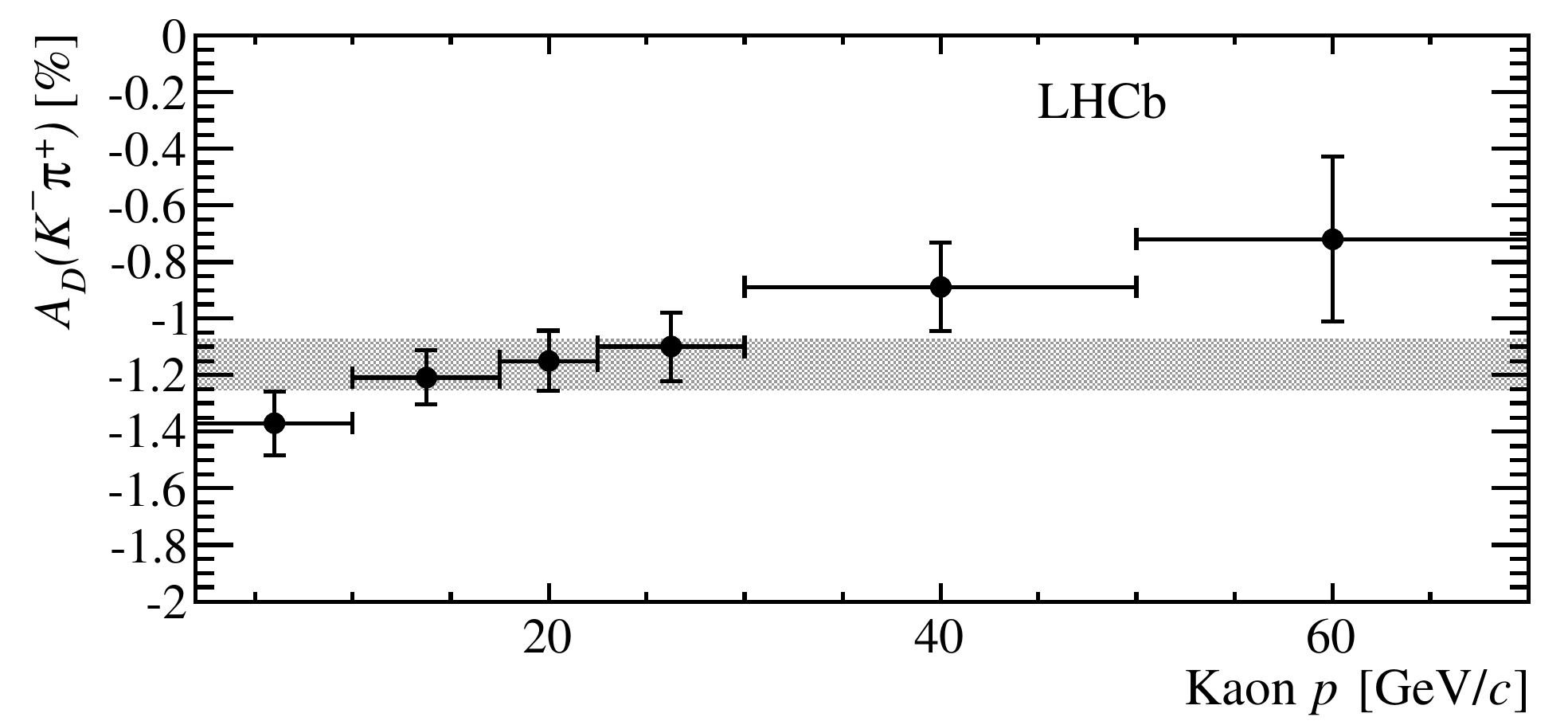}
  \end{center}
  \vspace{-0.5cm}
  \caption{Measured $\Km\pip$ detection asymmetry as a function of the kaon
    momentum. The shaded band indicates the average asymmetry integrated over
    the bins. There is a correlation between the data points due to
    the overlap between the \dkzpi samples used for each bin.}
  \label{fig:ADKpi}
\end{figure}

The raw asymmetries are determined with likelihood fits to the binned \Dz and
\Dp mass distributions using the mass models and event weights as described in
Sects.~\ref{sec:fitmodel} and \ref{sec:weighting}. The fits are done separately
for the 2011 and 2012 data sets and for the two magnet polarities. For each data
set the mean value of the raw asymmetry is the arithmetic average of the fit
results for the two magnet polarities. The final raw asymmetry is then the
statistically weighted average over the full data set. The derivation of the
$\Km\pip$ detection asymmetry using prompt \dkpipi and \dkzpi decays is shown in
Table~\ref{tab:kpi_asymmetry}. The measured asymmetry,
$\ADKpi=(-1.17\pm0.12)\%$, is dominated by the different interaction cross
sections of \Km and \Kp mesons in matter. Figure~\ref{fig:ADKpi} shows the
detection asymmetry as a function of the kaon momentum. As expected, the kaon
interaction asymmetry decreases with kaon momentum.

\renewcommand{\arraystretch}{1.2}
\begin{table}
  \begin{center}
    \caption{Asymmetries (in \%) entering the calculation of the $\Km\pip$
      detection asymmetry for the two magnet polarities, and for the mean
      value. The correction for the \Kz asymmetry is applied in the bottom
      row. The mean values in the last column are obtained first by taking the
      arithmetic average over the magnet polarities and then by taking the
      weighted averages of the 2011 and 2012 data sets. The uncertainties are
      statistical only.}
  \label{tab:kpi_asymmetry}
  \vspace{0.1cm}
  \begin{tabular}{l r@{$\pm$}lr@{$\pm$}lr@{$\pm$}l} \hline &
  \multicolumn{2}{c}{Magnet up} & \multicolumn{2}{c}{Magnet down} & 
  \multicolumn{2}{c}{Mean} \\ \hline
\ArawKpipi & $-1.969$&$0.033$ &  $-1.672$&$0.032$ & $-1.827$&$0.023$ \\ 
\ArawKzpi  & $-0.94$&$0.17$   &  $-0.51$&$0.16$   & $-0.71$&$0.12$   \\ \hline 
\ADKpi     & $-1.08$&$0.17$   &  $-1.22$&$0.16$   & $-1.17$&$0.12$   \\ \hline
\end{tabular}
\end{center}
\end{table}
\renewcommand{\arraystretch}{1.0}

\begin{figure}
  \begin{center}
    \includegraphics[width=0.49\textwidth]{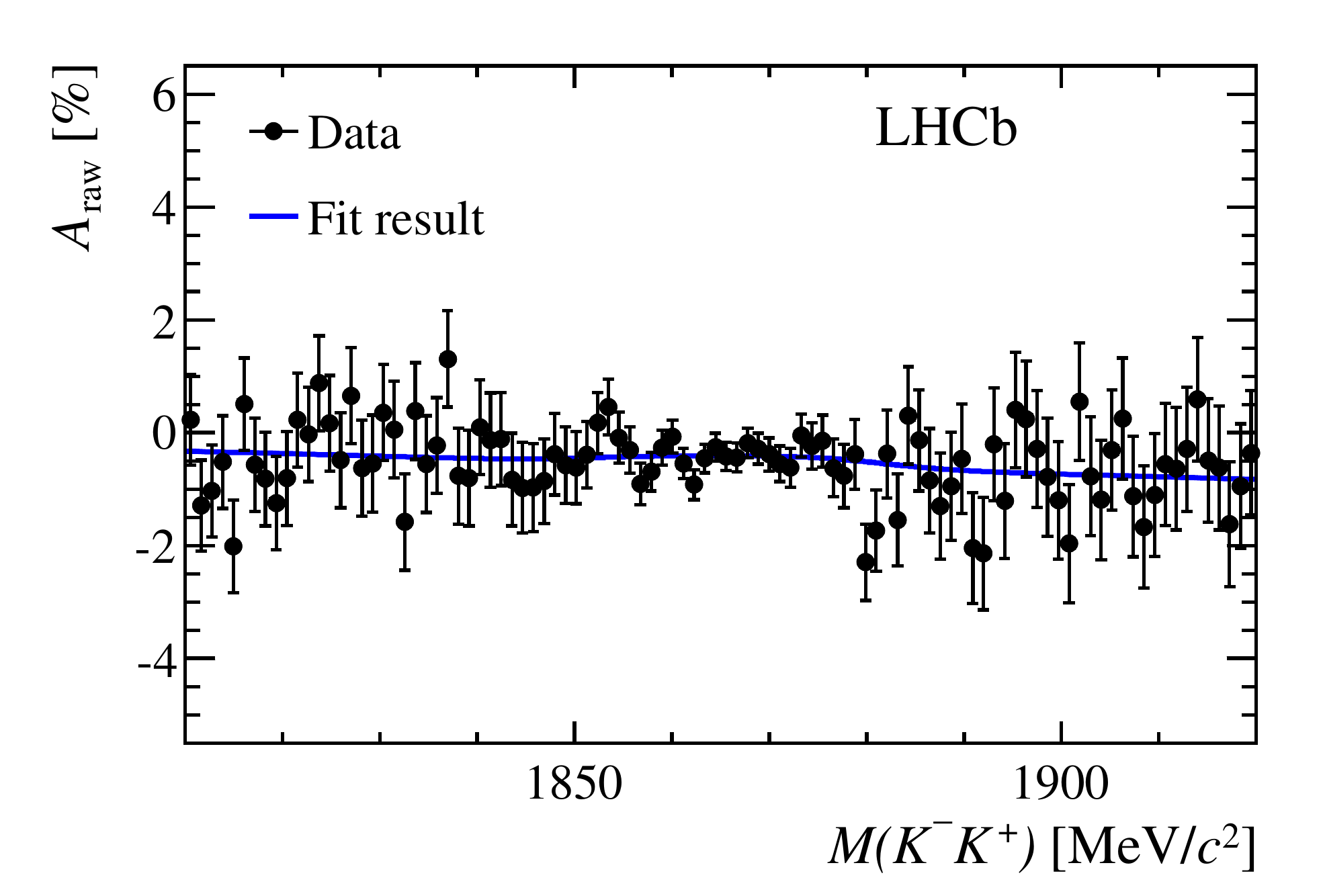}\put(-35,125){(a)}
    \includegraphics[width=0.49\textwidth]{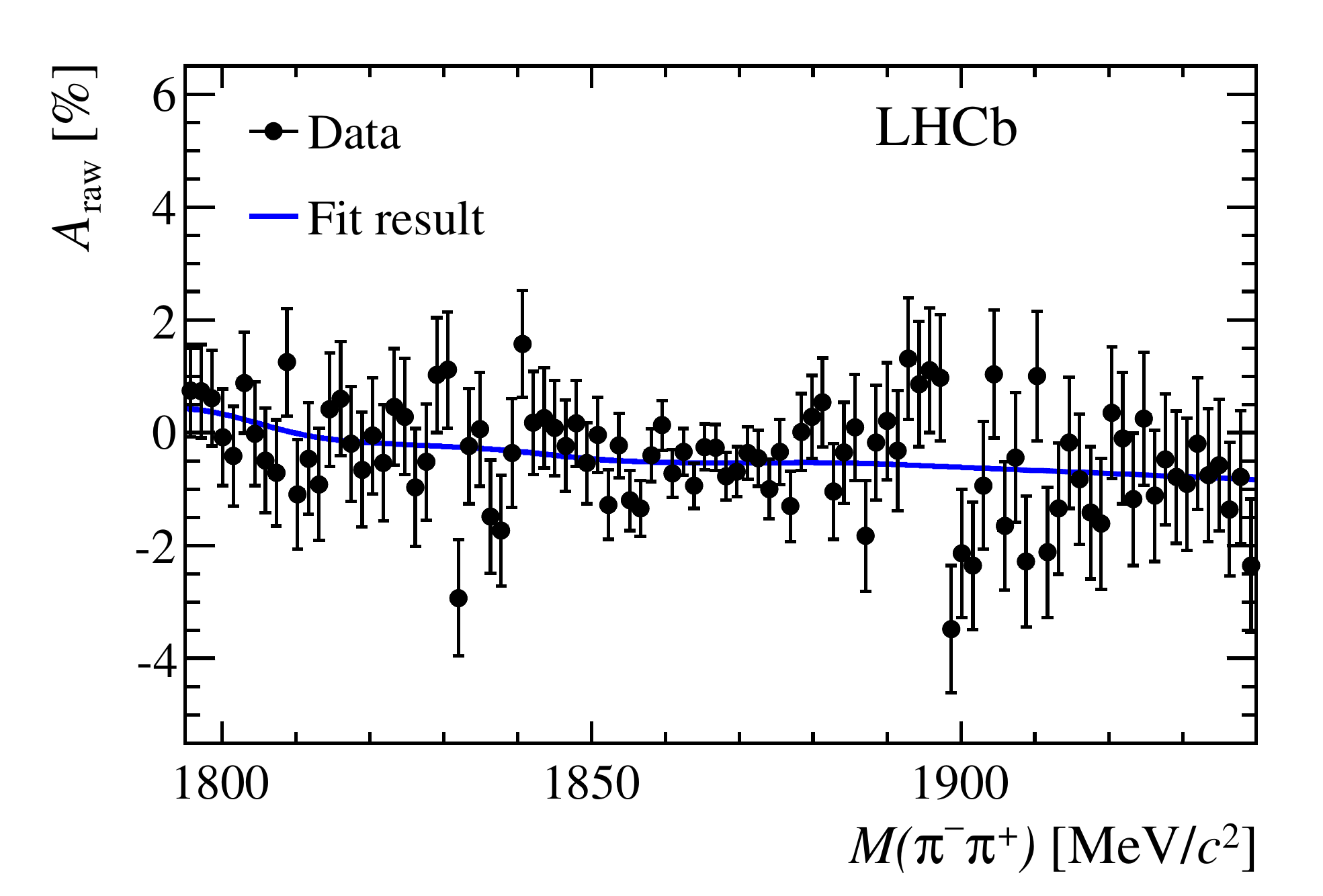}\put(-35,125){(b)}
  \end{center}
  \vspace{-0.5cm}
  \caption{Raw asymmetry, without background subtraction, as a function of the
    invariant mass for (a) the \dkk candidates and (b) the \dpipi candidates for
    the \DACP selection. The result from the fit is overlaid.}
  \label{fig:asym_vs_mass}
\end{figure}

For illustration, Fig.~\ref{fig:asym_vs_mass} shows the raw asymmetries for \dkk
and \dpipi candidates as functions of the invariant mass. The raw asymmetry in
both decay modes is slightly negative. The derivation of \DACP and \AcpKK from
the raw asymmetries are shown in Tables~\ref{tab:default_asymmetry} and
\ref{tab:acp_asymmetry}. There is a statistical correlation $\rho=\corrstat$
between the values of \DACP and \AcpKK as they both use candidates in the \dkk
sample.

\renewcommand{\arraystretch}{1.2}
\begin{table}
  \begin{center}
    \caption{Asymmetries (in \%) used in the calculation of \DACP for the two
      magnet polarities. The values for \DACP are corrected for the mistag
      probability. The mean values in the last column are obtained first by
      taking the arithmetic average over the magnet polarities and then by
      taking the weighted averages of the 2011 and 2012 data sets. The
      uncertainties are statistical only.}
  \label{tab:default_asymmetry}
  \vspace{0.1cm}
  \begin{tabular}{l r@{$\pm$}lr@{$\pm$}lr@{$\pm$}l} \hline &
  \multicolumn{2}{c}{Magnet up} & \multicolumn{2}{c}{Magnet down} & 
  \multicolumn{2}{c}{Mean} \\ \hline
\AKK     & $-0.46$&$0.11$ & $-0.43$&$0.11$ & $-0.44$&$0.08$ \\ 
\Apipi   & $-0.45$&$0.20$ & $-0.66$&$0.19$ & $-0.58$&$0.14$ \\  \hline 
\DACP    & $-0.01$&$0.23$ & $+0.24$&$0.22$ & $\DACPval$&$\DACPstat$ \\ \hline 
\end{tabular}
\end{center}
\end{table}
\renewcommand{\arraystretch}{1.0}

\renewcommand{\arraystretch}{1.2}
\begin{table}
  \begin{center}
    \caption{Asymmetries (in \%) used in the calculation of \AcpKK for the two
      magnet polarities. The values for \AcpKK are corrected for the mistag
      probability. The mean values in the last column are obtained first by
      taking the arithmetic average over the magnet polarities and then by
      taking the weighted averages of the 2011 and 2012 data sets. The
      uncertainties are statistical only.}
  \label{tab:acp_asymmetry}
  \vspace{0.1cm}
  \begin{tabular}{l r@{$\pm$}lr@{$\pm$}lr@{$\pm$}l}\hline &
  \multicolumn{2}{c}{Magnet up} & \multicolumn{2}{c}{Magnet down} & 
  \multicolumn{2}{c}{Mean} \\ \hline
\AKK     & $-0.45$&$0.12$ & $-0.41$&$0.12$ & $-0.43$&$0.08$ \\ 
\ArawKpi & $-1.41$&$0.05$ & $-1.59$&$0.05$ & $-1.51$&$0.04$ \\ 
\ADKpi   & $-1.08$&$0.17$ & $-1.22$&$0.16$ & $-1.17$&$0.12$ \\ \hline 
\AcpKK   & $-0.09$&$0.21$ & $-0.01$&$0.21$ & $\AcpKKval$&$\AcpKKstat$ \\ \hline
\end{tabular}
\end{center}
\end{table}
\renewcommand{\arraystretch}{1.0}

\section{Systematic uncertainties}
\label{sec:systematics}

Systematic shifts in the observed \CP asymmetries can arise from
non-cancellation of production and detection asymmetries, misreconstruction
of the final state, and imperfect modelling of the background. The
contributions to the systematic uncertainties in \DACP and \AcpKK are described
below.

The fractions of \Bd and \Bp decays in the three semileptonic \B samples can be
slightly different. Assuming that there is a difference in the \Bd and \Bp
production asymmetries, a residual production asymmetry can remain in \DACP and
\AcpKK. As in the previous publication~\cite{LHCb-PAPER-2013-003}, a systematic
uncertainty of $0.02\%$ is assigned to both \DACP and \AcpKK.  Due to \Bz
oscillations, the observed \B production asymmetry depends on the decay-time
acceptance of the reconstructed \B meson, which is slightly different for the
three decay modes. This produces a systematic uncertainty of $0.02\%$ for both
\DACP and \AcpKK, similar to the one found
previously~\cite{LHCb-PAPER-2013-003}.

The weighting procedure almost equalises the \pt and $\eta$ distributions of the
particles, but small differences remain. When also weighting for different
azimuthal angle distributions of the final state particles, the change in both
\DACP and \AcpKK is negligible. Slightly larger shifts are seen when increasing
(decreasing) the number of bins used in each kinematic variable from 20 to 25
(15) or when changing the \Dz mass range. The maximum shift, $0.02\%$ for \DACP
and $0.05\%$ for \AcpKK, is taken as a systematic uncertainty. Finally, the
cancellation of the production and detection asymmetries is tested by randomly
assigning the charge of the muon or charged \D meson in real data, depending on
the \pt of the particles. The \B and \Dp production asymmetries and the
$\mu^{\pm}$, \Kpm and \pipm detection asymmetries that are simulated in this way
are motivated by the small \pt-dependences observed in data. No shift is seen in
the value of \DACP, while a small shift of $0.03\%$ is observed in the value of
\AcpKK. This shift is propagated as part of the uncertainty due the weighting
procedure.

The sensitivity of the results to the signal and background models is determined
by varying the signal and background functions. The alternative signal functions
are a Johnson $S_U$ distribution~\cite{Johnson:1949zj}, a single Gaussian, and a
double Gaussian function. The alternative background function is a second-order
polynomial. Furthermore, the effect of using different mass binning and fit
range, and the effect of constraining the asymmetry of the \dkpi background in
the \dpipi decay to the observed asymmetry, are considered. The maximum
variations from the default fit for each decay mode are added in quadrature to
determine the systematic uncertainty for \DACP ($0.06\%$) and for \AcpKK
($0.06\%$).

In the default fit, the background can vary freely with an overall asymmetry and
different slope parameters for each tag. Nevertheless, background contributions
from different origins can have different shapes and asymmetries. Such an effect
is expected to be largest in the \dkk decay, due to possible contributions from
other charm decays, and is studied by generating pseudoexperiments with
different background compositions in the two Cabibbo-suppressed decays. Three
types of background shapes are simulated: an exponential function to describe
the combinatorial background observed in data, another exponential function with
a different slope inspired by partially-reconstructed background from simulated
\dkpipi decays, and a linear shape inspired by partially-reconstructed
background from simulated $\Lc\to\Pp\Km\pip$ decays. The asymmetries in the
background are varied by up to $\pm3\%$. Such large asymmetries are incompatible
with the asymmetries observed in the background and therefore constitute an
upper bound on the magnitude of any possible effect. The largest bias in the raw
asymmetry ($0.03\%$) is propagated as a systematic uncertainty for \DACP and
\AcpKK.

The systematic shift in the raw asymmetries when removing multiple candidates is
below $0.005\%$ and therefore neglected. Higher-order corrections to
Eq.~(\ref{eq:Araw}) are at the $10^{-6}$ level and are neglected as well. The
systematic uncertainty from the neutral kaon asymmetry ($0.01\%$) is taken from
Sect.~\ref{sec:KzAsymmetry} and the systematic uncertainty from wrong
combinations of muons and \Dz mesons is taken from Sect.~\ref{sec:mistag}. All
systematic uncertainties are summarised in Table~\ref{tab:syst} for \DACP and
\AcpKK. The correlation coefficient between the total systematic uncertainties
is $\rho=0.40$.

\begin{table}
  \begin{center}
    \caption {\small Contributions to
    the systematic uncertainty of \DACP and \AcpKK.}
    \label{tab:syst}
    \vspace{0.1cm}
    \begin{tabular}{l c c} \hline
  Source of uncertainty                      & \DACP     & \AcpKK   \\ \hline
  {Production asymmetry:}                    &           &          \\
  ~~~~Difference in \bquark-hadron mixture   & $0.02\%$  & $0.02\%$ \\
  ~~~~Difference in \B decay time acceptance & $0.02\%$  & $0.02\%$ \\
  {Production and detection asymmetry:}      &           &          \\
  ~~~~Different weighting                    & $0.02\%$  & $0.05\%$ \\
  ~~~~Non-cancellation                       &    -      & $0.03\%$ \\
  ~~~~Neutral kaon asymmetry                 &    -      & $0.01\%$ \\
  {Background from real \Dz mesons:}         &           &          \\
  ~~~~Mistag asymmetry                       & $0.03\%$  & $0.03\%$ \\ 
  {Background from fake \Dz mesons:}         &           &          \\
  ~~~~\Dz mass fit model                     & $0.06\%$  & $0.06\%$ \\ 
  ~~~~Wrong background modelling             & $0.03\%$  & $0.03\%$ \\
  \hline
  {Quadratic sum}                      &$\DACPsyst\%$&$\AcpKKsyst\%$\\
  \hline
    \end{tabular}
  \end{center}
\end{table}

\section{Consistency checks}
\label{sec:crosschecks}

As a consistency check, the raw asymmetries in the \dkk and \dpipi samples and
\DACP are determined as functions of the impact parameter of the \Dz trajectory
with respect to the primary vertex, the flight distance of the \B candidate, the
angle between the directions of the muon and the \Dz decay products, the muon
and \Dz kinematic variables, the reconstructed $\Dz\mu$ invariant mass, the
multiplicity of tracks and primary vertices in the event, the particle
identification requirement on the kaons, and the selected trigger lines. No
significant dependence is observed on any of these variables.
Another test is made by including \Dz candidates with negative decay times. In
particular in the \dpipi decay, there is more background at low \Dz decay
times. Enhancing this type of background by including negative decay time
candidates does not change the values for \DACP or \AcpKK.

\begin{figure}
  \begin{center}
    \includegraphics[width=0.49\textwidth]{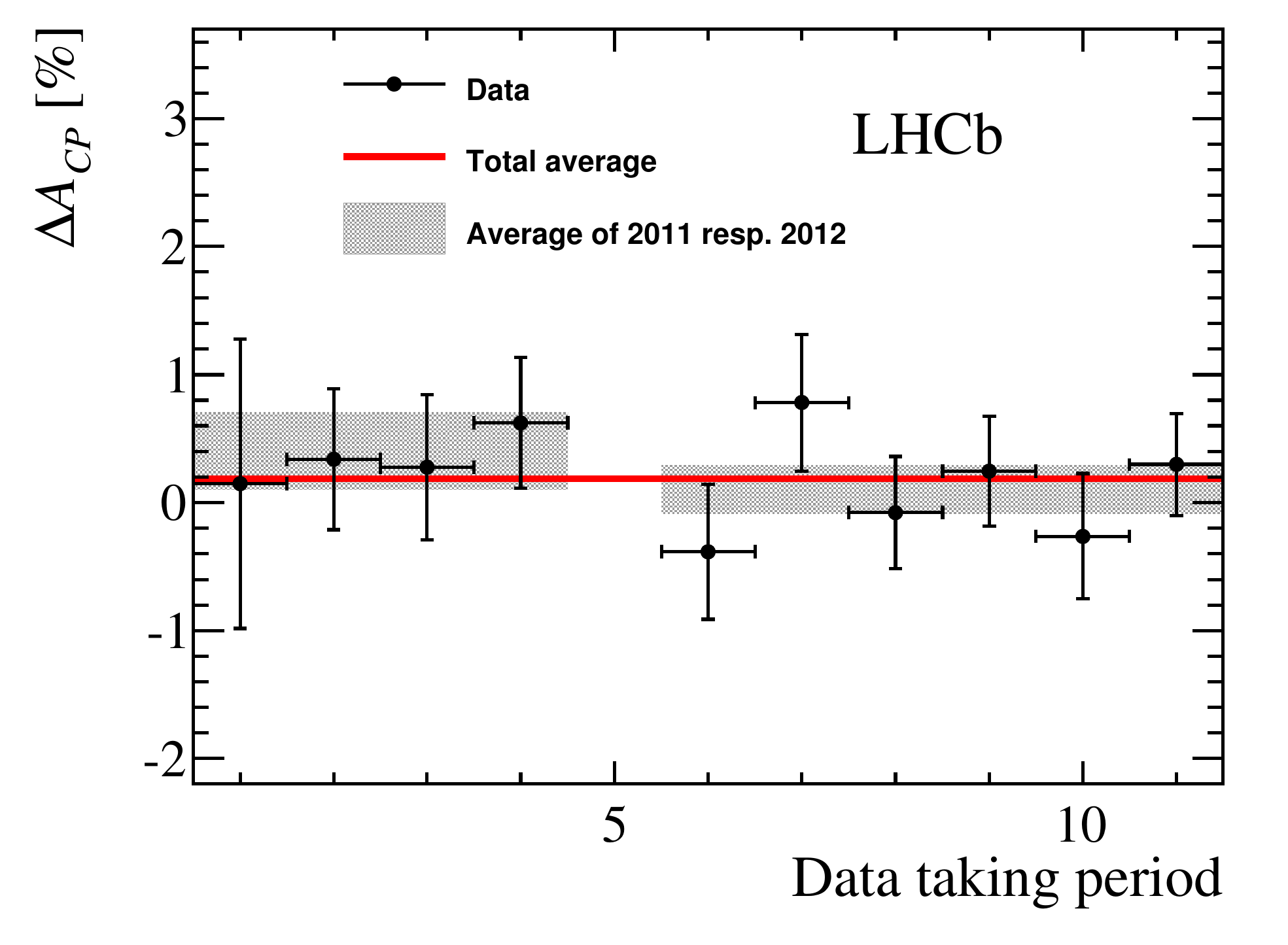}\put(-40,137){(a)}
    \includegraphics[width=0.49\textwidth]{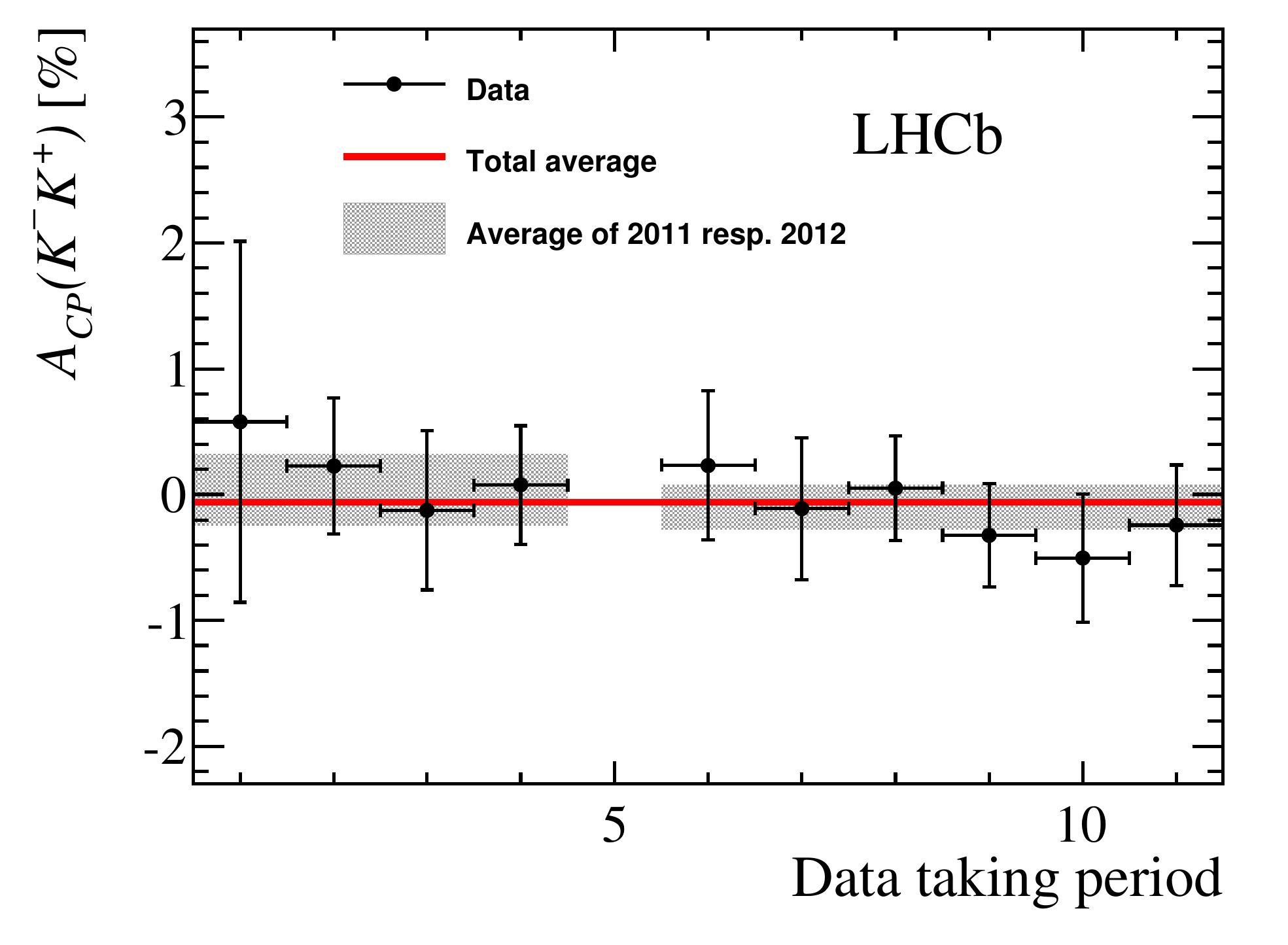}\put(-40,137){(b)}
  \end{center}
  \vspace{-0.5cm}
  \caption{(a) \DACP and (b) \AcpKK as a function of the data taking period. The
    2011 data are divided into four periods and the 2012 data into six
    periods. The error bars indicate the statistical uncertainty, the shaded
    bands show the averages for 2011 and 2012, and the (red) line shows the
    overall \CP asymmetry.}
  \label{fig:runPeriod}
\end{figure}

During periods without data taking, interventions on the detector and on the
trigger change the alignment and data acquisition conditions. This could induce
time-varying detection asymmetries. The final results should not be sensitive to
such variations as data are calibrated with control samples collected in the
same data-taking period. Nevertheless, any residual detector asymmetry would
manifest itself as variations in time of the measured \CP
asymmetries. Figure~\ref{fig:runPeriod} shows \DACP and \AcpKK versus data
taking period. These periods are separated by interruptions in data taking, and
within each period the magnetic field is reversed at least once. No dependence
of the obtained \CP asymmetries on the data taking period is observed. The
average values of \DACP are $(+0.33\pm0.30\stat)\%$ for the 2011 data and
$(+0.06\pm0.19\stat)\%$ for the 2012 data. The value for the 2011 data is
slightly lower compared to the previous analysis, which is attributed to the
non-overlapping data samples, due to differences in the selection and in the
calibration of the detector. The main shift is due to the looser particle
identification requirements on the kaons in this analysis. Such a shift is not
seen in the 2012 data. The average values of \AcpKK are $(+0.04\pm0.28\stat)\%$
for the 2011 data and $(-0.10\pm0.18\stat)\%$ for the 2012 data.

\section{Conclusion}
\label{sec:conclusion}

The difference in \CP asymmetries between the \dkk and \dpipi decay channels and
the \CP asymmetry in the \dkk channel are measured using muon-tagged \Dz decays
in the 3\invfb data set to be
\begin{align}
  \DACP  &= (\DACPval \pm \DACPstat\stat \pm \DACPsyst\syst)\%
  \ , \nonumber \\
  \AcpKK &= (\AcpKKval\pm \AcpKKstat\stat \pm \AcpKKsyst\syst) \%  \ ,\nonumber 
\end{align}
where the total correlation coefficient, including statistical and
systematic components, is $\rho=\corrtot$. By combining the above
results, the \CP asymmetry in the \dpipi decay is found to be
\begin{equation}
  \Acppipi=(\Acppipival\pm\Acppipistat\stat\pm\Acppipisyst\syst)\% \ . \nonumber
\end{equation}
These results are obtained assuming that there is no \CP violation in \Dz mixing
and no direct \CP violation in the Cabibbo-favoured \dkpi, \dkpipi and \dkzpi
decay modes. The measurement of \DACP supersedes the previously reported
result~\cite{LHCb-PAPER-2013-003}.  Our results show that there is no
significant \CP violation in the singly Cabibbo-suppressed
$\Dz\to\Km\Kp,\pim\pip$ decays at the level of $10^{-3}$. These results
constitute the most precise measurements of time-integrated \CP asymmetries
\AcpKK and \Acppipi from a single experiment to date.

\section*{Acknowledgements}

\noindent We express our gratitude to our colleagues in the CERN accelerator
departments for the excellent performance of the LHC. We thank the technical and
administrative staff at the LHCb institutes. We acknowledge support from CERN
and from the national agencies: CAPES, CNPq, FAPERJ and FINEP (Brazil); NSFC
(China); CNRS/IN2P3 and Region Auvergne (France); BMBF, DFG, HGF and MPG
(Germany); SFI (Ireland); INFN (Italy); FOM and NWO (The Netherlands); SCSR
(Poland); MEN/IFA (Romania); MinES, Rosatom, RFBR and NRC ``Kurchatov
Institute'' (Russia); MinECo, XuntaGal and GENCAT (Spain); SNSF and SER
(Switzerland); NASU (Ukraine); STFC and the Royal Society (United Kingdom); NSF
(USA). We also acknowledge the support received from EPLANET, Marie Curie
Actions and the ERC under FP7.  The Tier1 computing centres are supported by
IN2P3 (France), KIT and BMBF (Germany), INFN (Italy), NWO and SURF (The
Netherlands), PIC (Spain), GridPP (United Kingdom).  We are indebted to the
communities behind the multiple open source software packages on which we
depend.  We are also thankful for the computing resources and the access to
software R\&D tools provided by Yandex LLC (Russia).

\addcontentsline{toc}{section}{References}
\setboolean{inbibliography}{true}
\bibliographystyle{LHCb}
\bibliography{main,LHCb-PAPER,LHCb-CONF,LHCb-DP}


\end{document}